\documentclass[usenatbib,usegraphicx]{mn2e}
\pdfoutput=1
\usepackage{bm}
\usepackage{amssymb}
\usepackage[fleqn]{amsmath}
\usepackage{charter}
\usepackage{mathdesign}
\usepackage{url}
\usepackage{color}
\usepackage[usenames,dvipsnames]{xcolor}
\usepackage[hyperfootnotes, hyperfigures,bookmarks, pdftex]{hyperref}
\hypersetup{bookmarksnumbered, bookmarksopen, colorlinks, citecolor=blue, pdfduplex=DuplexFlipLongEdge}
\hypersetup{pdftitle=CDI Morphology and Energetics, pdfauthor=Sean M. O'Neill, pdfsubject=Astrophysics}
\usepackage{bookmark}
\providecommand{\adsurl}[1]{\href{#1}{ADS}}

\newcommand{\gtorder}{\mathrel{\raise.3ex\hbox{$>$}\mkern-14mu
             \lower0.6ex\hbox{$\sim$}}}
\newcommand{\ltorder}{\mathrel{\raise.3ex\hbox{$<$}\mkern-14mu
             \lower0.6ex\hbox{$\sim$}}}

\newcommand\eg{{e.g.,~}}
\newcommand\ie{{i.e.,~}}

\title[CDI Morphology and Energetics]{Local Simulations of Instabilities in Relativistic Jets I: \\ Morphology and Energetics of the Current-Driven Instability}
\author[O'Neill et al.]{Sean M. O'Neill$^{1 \ast}$, Kris Beckwith$^{1,2 \star}$, Mitchell C. Begelman$^{1,3 \dagger}$ \\
$^1${JILA, University of Colorado and National Institute of Standards and Technology, 440 UCB, Boulder, CO 80309-0440, USA}  \\
$^2${Tech-X Corporation, 5621 Arapahoe Ave. Suite A, Boulder, CO 80303, USA} \\
$^3${Department of Astrophysical and Planetary Sciences, University of Colorado, 391 UCB, Boulder, CO 80309-0391, USA} \\
Email: $^\ast$sean.m.oneill@jila.colorado.edu,
$^\star$kris.beckwith@jila.colorado.edu,
$^\dagger$mitch@jila.colorado.edu}

\begin{document}

\label{firstpage}

\maketitle

\begin{abstract}
We present the results of a numerical investigation of current-driven instability in magnetized jets.
Utilizing the well-tested, relativistic magnetohydrodynamic code {\tt Athena}, we construct an ensemble of local, co-moving plasma columns in which initial radial force balance is achieved through various combinations of magnetic, pressure, and rotational forces.
We then examine the resulting flow morphologies and energetics to determine the degree to which these systems become disrupted, the amount of kinetic energy amplification attained, and the non-linear saturation behaviors.
Our most significant finding is that the details of initial force balance have a pronounced effect on the resulting flow morphology.
Models in which the initial magnetic field is force-free deform, but do not become disrupted.
Systems that achieve initial equilibrium by balancing pressure gradients and/or rotation against magnetic forces, however, tend to shred, mix, and develop turbulence.
In all cases, the linear growth of current-driven instabilities is well-represented by analytic models.
CDI-driven kinetic energy amplification is slower and saturates at a lower value in force-free models than in those that feature pressure gradients and/or rotation.
In rotating columns, we find that magnetized regions undergoing rotational shear are driven toward equipartition between kinetic and magnetic energies.
We show that these results are applicable for a large variety of physical parameters, but we caution that algorithmic decisions (such as choice of Riemann solver) can affect the evolution of these systems more than physically motivated parameters.
\end{abstract}


\section{Introduction}
Collimated, magnetized flows are ubiquitous in astrophysics, but detailed studies of the stability and evolution of such systems have focused primarily on the shear-driven Kelvin-Helmholtz instability (KHI, see \eg \citealt{2006AIPC..856...57H} or \citealt{salvesenetal2012} for review), where magnetic forces are often relegated to a supporting role.
In fact, much of the pioneering work on the intrinsically magnetized family of current-driven instabilities (CDI) was conducted by the plasma physics community with the goal of understanding confined plasma columns in the context of Tokamak-type fusion experiments.
Nevertheless, there are many obvious astrophysical circumstances where CDI have the potential to play a significant role.

Jets emerging from accretion onto black holes, for example, are very likely to be magnetically launched (\eg \citealt{1976Natur.262..649L}, \citealt{1976MNRAS.176..465B}, \citealt{1982MNRAS.199..883B}) and subsequently magnetically dominated near their points of origin.
Such flows will generally be subject to CDI, but the ramifications of instability in this context remain unclear, particularly when the jet is bounded by a strong shear layer.
Even if a shear layer confines lateral jet expansion and disruption caused by CDI, one could still imagine CDI-driven dissipation resulting in variability signatures, such as the very rapid TeV variability recently observed in blazars Mrk 501 and PKS 2155-304 \citep{2007ApJ...664L..71A,2007ApJ...669..862A}.
At greater distances from their origin, jets from active galactic nuclei (AGN) are also inferred observationally to transition to a kinetic-energy dominated flow \citep{2005ApJ...625...72S}, a process which itself may very well involve CDI-driven dissipation.
Even in such a weakly magnetized regime, however, CDI have the potential to grow and subsequently to disrupt otherwise laminar flows.

This is not an issue exclusively relevant to ultra-relativistic jets.
In the context of filled supernova remnants (\ie plerions), notably including the Crab Nebula, CDI have been identified as a possible means of breaking axisymmetry in the outflowing pulsar-driven wind \citep{ 1998ApJ...493..291B}.
If present, this process could obviate commonly made geometric assumptions that constrain the magnetization of the flow at the wind shock to be significantly less than unity (\eg \citealt{1974MNRAS.167....1R}). 
Known as the ``$\sigma$ problem,'' this inference has long been at odds with the fact that the wind is likely to be magnetically dominated near the pulsar \citep{1970ApJ...160..971G} while simultaneously unlikely to convert enough of its magnetic energy before reaching the shock (\eg \citealt{1994ApJ...426..269B}), but CDI-driven asymmetry may provide a solution to this seeming paradox.
Furthermore, it is possible that dissipation events facilitated by instabilities such as the CDI could produce the recently observed, rapid, gamma-ray flares observed in the Crab Nebula by AGILE \citep{2011Sci...331..736T} and {\it Fermi} \citep{2011Sci...331..739A}.

There has been much discussion over the years in both the plasma physics and astrophysics communities on the linear development of the CDI under various physical conditions, with much of the pioneering work emerging in the latter half of the last century.
\citet{1951PhRv...83..307L}, for example, identified a criterion describing the ratio of poloidal to toroidal field that would cause a magnetized cylinder to become unstable.
\citet{1954RSPSA.223..348K}  further demonstrated analytically that a plasma column threaded by a purely toroidal field (known in fusion literature as a ``Z-pinch'') is characteristically unstable.
This work was later generalized by \citet{1957PPSB...70...31T} to multiple mode numbers.
Additionally, \citet{1966RvPP....2..153K} identified a criterion for which the axisymmetric m=0 mode of CDI (commonly referred to as the ``sausage instability'') is stable.
Useful overviews of the basic physics of CDI, including distinctions between the internal (fixed-boundary) and external (free-boundary) kink modes and discussions of cylindrical column stability, have appeared in a number of textbooks, including \citet{1978mit..book.....B} and \citet{2004prma.book.....G}, and a review article by \citet{2004RvMP...76.1071B}.

In the context of astrophysics, stability analyses have been conducted in the non-relativistic limit by a large number of groups \citep{1983ApJ...269..500C,1989PhFlB...1..923P,1990PhFlB...2..828C,1992A&A...256..354A,1996A&A...314..995A,2000A&A...355..818A,2000A&A...363.1166K,2011A&A...525A.100B} under various sets of assumptions.
Research investigating stability in the relativistic regime has been undertaken primarily in the force-free limit \citep{1994MNRAS.267..629I,1996MNRAS.281....1I,1999MNRAS.308.1006L,2001PhRvD..64l3003T,2009ApJ...697.1681N}, with the exception of \citet{1998ApJ...493..291B}, which did not make the force-free assumption.
This latter work will form the primary point of comparison between analytic theory and those of our simulations that include a pressure gradient that initially balances the magnetic forces.

Magnetohydrodynamic simulations of CDI have been conducted in a number of different contexts.
In the non-relativistic regime, \citet{2000A&A...355.1201L},  \citet{2002ApJ...580..800B}, \citet{2004ApJ...617..123N},  \citet{2006ApJ...643...92L},  \citet{2007ApJ...656..721N},  \citet{2008ApJ...686..843N}, \citet{2008A&A...492..621M}, and \citet{2009NuPhS.190...88I} have explored both local and global models of CDI evolution.
Relativistic numerical investigations of CDI, however, have thus far been conducted by a relatively small number of groups.
In a recent series of papers, \citet{2009ApJ...700..684M} and \citet{2011ApJ...728...90M} have run and analyzed local simulations of CDI both generally and in the specific context of 
pulsar wind nebulae.
These models, simulated in a reference frame that is co-moving with the bulk jet flow, were designed to explore the evolution of CDI under the most idealized circumstances.
They further extended this work to local models bounded by a shear layer to explore interactions between CDI and the KHI \citep{2011ApJ...734...19M}, providing an extension into the relativistic limit of an earlier investigation by \citet{2002ApJ...580..800B}.
Additionally, \citet{2009MNRAS.394L.126M} have conducted a set of global general relativistic simulations, finding that their jets develop only limited substructure resulting from the action of CDI, although \citet{2010MNRAS.402....7M} identify ``jet wiggling'' in their global special relativistic jets that they do attribute to CDI.

Our goal in this work is to construct a large number of idealized jet configurations to determine various circumstances under which CDI develop and to examine the resulting flow morphologies and energetics.
Adopting a similar approach to that of \citet{2009ApJ...700..684M}, we will exclusively consider {\it local} models in which the observer is co-moving with the bulk flow, far removed from any jet boundaries.
Unlike previous work, however, we will explore several different arrangements of the initial forces, considering systems in which the magnetic fields are force-free and those in which the $\mathbf{J \times B}$ force is balanced by pressure gradients and/or rotation.
The results of these simulations will be compared both to analytic estimates of CDI linear growth and to the results of previously conducted numerical simulations.
Additionally, we will determine whether special relativity plays an important role in the evolution of these systems even in a reference frame for which the bulk velocity has been transformed away.
Section \ref{sec:numerics} outlines our numerical methods and the initial conditions of our various models of magnetized plasma columns.
Section \ref{sec:results} describes the results of our simulations, in terms of the simulated column morphologies, energetics, and how these quantities depend on various physical and numerical inputs.
Finally, we discuss the implications of our models and present our conclusions in Section \ref{sec:conclusions}.

\section{Numerical Details}
\label{sec:numerics}

The calculations presented in this work were conducted using {\tt Athena}, a second-order accurate Godunov flux-conservative code for solving the equations of magnetohydrodynamics (MHD)\footnotemark.
\footnotetext{The {\tt Athena} code and a repository of test problems are maintained online at \url{https://trac.princeton.edu/Athena/}.}
The basic algorithms implemented in {\tt Athena} are described by \citet{2005JCoPh.205..509G,2008JCoPh.227.4123G} with further details (implementation and multi-dimensional tests) given in \citet{2008ApJS..178..137S}.
Specifically, we utilize the dimensionally unsplit MUSCL-Hancock integrator (``Van-Leer'') described by \citet{2009NewA...14..139S} combined with the constrained transport (CT) method of \citet{2005JCoPh.205..509G,2008JCoPh.227.4123G} to maintain the divergence-less nature of the magnetic field in multi-dimensions, extended to relativistic MHD (RMHD) as described by \citet{2011ApJS..193....6B}.

The RMHD module within the {\tt Athena} code implements a variety of Riemann solvers and spatial reconstruction methods \cite[as described in][]{2011ApJS..193....6B}. 
\cite{2011ApJS..193....6B} demonstrate that the choice of Riemann solver can play an important role in determining the \emph{spectral} accuracy of a numerical scheme. 
Even if a given scheme is stable and consistent, so that solutions converge to a weak solution of the conservation law, the solution itself may not be \emph{unique}. 
That is, algorithms with different spectral performance may converge to \emph{different} weak solutions of the conservation law. 
As a result, it is important to assess the impact of different algorithmic choices (such as Riemann solver and spatial reconstruction method) on the outcome of simulations of CDI. 
The impact of these choices, along with that of numerical resolution, on the development of CDI is examined in detail \S\ref{sec:spectral}.

\subsection{Initial Conditions}
\label{sec:initialconditions}
In this work, we study the development of CDI in a set of idealized systems in order that we may understand the physics behind the phenomena that develop.
To do so, we conduct three basic categories of simulation for which initial force balance is achieved in three different ways.
In the Newtonian approximation and in the absence of an external force, hydromagnetic force balance of an ideal MHD fluid can be expressed as
\begin{equation}
0=-(\mathbf{v \cdot \nabla)v} -\frac{\mathbf{\nabla} p}{\rho} + \frac{1}{\rho}(\mathbf{J \times B}),
\end{equation}
where $\mathbf{v}$ is the velocity field, $p$ is the gas pressure, $\rho$ is the density, $\mathbf{B}$ is the magnetic field, and $\mathbf{J} = \mathbf{\nabla \times B}$ is the current density. In constructing this force balance equation, we have taken units where $G=M=c=4\pi=1$.
If we further assume that the velocity field is restricted to a rotational field of the form $\mathbf{v} = v_{\phi}(r) \bm{\hat{\phi}}$ and consider only radial force balance, this expression simplifies to
\begin{equation}
0=\frac{v_{\phi}^2}{r} -\frac{(\mathbf{\nabla} p)_r}{\rho} + \frac{1}{\rho}(\mathbf{J \times B})_r,
\label{eq:force}
\end{equation}
where the first term is the (exclusively non-negative) centrifugal acceleration, the second is the pressure gradient, and the third reflects magnetic pressure and/or tension.
In our first category of simulation, which we label force-free (FF), each of these terms is independently set to zero, meaning that the system initially features uniform pressure, no rotation, and a non-zero magnetic field.
In the second type (PB) of simulation, the columns are initialized such that the pressure gradient and magnetic forces balance one another in the absence of rotation. 
In the third type (RPB), initial force balance is achieved through a combination of non-zero rotational, pressure, and magnetic forces.
Table 1 provides a listing of all production-level simulations, and we will now explain the different types of initial conditions employed in each. In all of these simulations, we assume an adiabatic equation of state with a constant $\gamma=5/3$, except as noted.

\begin{table}
\begin{center}
\caption{Summary of Jet Column Simulations}
\label{sims}
\begin{tabular}{ @{}l|cccc}
\hline
{{\bf ID}} & {\bf FIELD} & ${\bm \beta}$ & {\bf T ($\tau_{\rm A}$)} & {\bf OTHER NOTES} \\ \hline
\hline
{\bf PB} & sinusoidal & 0.3 & 20 & \\ \hline
PB-$\beta$3 & sinusoidal & 3 & 20 & \\ \hline
PB-$\beta$30 & sinusoidal & 30 & 11 & \\ \hline
PB-kom & Komissarov & 0.3 & 20 & \\ \hline
PB-bz1 & sinusoidal & 0.3 & 20 & $B_z = -B_\phi$ \\ \hline
PB-bz0.2 & sinusoidal & 0.3 & 20 & $B_z = -0.2 B_\phi$ \\ \hline
PB-g1.33 & sinusoidal & 0.3 & 20 & $\gamma=4/3$\\ \hline
PB-newt & sinusoidal & 0.3 & 20 & Newt. \\ \hline
PB-vrand & sinusoidal & 0.3 & 20 & v$_{\rm kick}$ randomized \\ \hline
\hline
{\bf RPB} & sinusoidal & 0.3 & 69 & Newt. \\ \hline
RPB-slow & sinusoidal & 0.3 & 49 & Newt., decreased $v_\phi$ \\  \hline
RPB-xrot & sinusoidal & 0.3 & 52 & Newt., extended $v_\phi$ \\ \hline
\hline
{\bf FF} & force-free & 0.25 & 600 & \\ \hline
FF-hot & force-free & 0.25 & 600 & $p \sim 10 \rho$, $\gamma=4/3$\\\hline
FF-v6 & force-free & 0.25 & 600 & v$_{\rm kick}=0.06 v_{\rm A}$\\ \hline
FF-v25 & force-free & 0.25 & 600 & v$_{\rm kick}=0.25 v_{\rm A}$\\ \hline
FF-newt & force-free & 0.25 & 600 & Newt. \\ \hline
FF-vrand & force-free & 0.25 & 600 & v$_{\rm kick}$ randomized \\ \hline
\hline
\vspace{-1mm}
\end{tabular}
\end{center}
\medskip Here, FIELD is magnetic field configuration, as described in Section \ref{sec:initialconditions}.  $\beta$ is minimum value of $\beta \equiv 2 p/B^2$ on the grid.  $T$ is simulation duration in units of $\tau_{\rm A} \equiv r_c/v_{\rm A}$, where $v_{\rm A}$ is the initial maximum Alfv\'en speed on the grid and $r_c$ is the initial column radius.  Newt. refers to Newtonian physics (as opposed to SR).
\end{table}

The magnetic fields in our models, all of which can be represented in cylindrical coordinates as $\mathbf{B}=B_{\phi}\mathbf{\bm{\hat{\phi}}}+B_z\mathbf{\hat{z}}$, fall into three generic categories that we refer to in Table 1 as sinusoidal, Komissarov, and force-free.  For the sinusoidal field configuration,
\begin{equation}
B_{\phi}(r)   = 
\left\{
\begin{array}{c}
\begin{aligned}
& 0.5 B_0 (1-\cos[2\pi r/r_{\rm c}]) &~~~~~& r \le r_{\rm c} \\
& 0 &~~~~~& r > r_c
\end{aligned}
\end{array}
\right.,
\end{equation}
where $B_0$ is a fiducial field strength, $r$ is the radial cylindrical coordinate, and $r_c$ is the characteristic column radius.
Typically, $B_z=0$ for sinusoidal fields, corresponding to a magnetic pitch angle $P \equiv rB_z / B_\phi = 0$.
The sole exceptions are the PB-bz1 and PB-bz0.2 models, where a poloidal field is present such that $B_z(r) \propto B_{\phi}(r)$, with the constants of proportionality provided for each model in Table 1.
What we refer to as the Komissarov-type field (\eg \citealt{1999MNRAS.308.1069K}) is a purely toroidal field in which 
\begin{equation}
B_{\phi}(r)   = 
\left\{
\begin{array}{c}
\begin{aligned}
& B_{0}  (r/r_{\rm c}) &~~~~~& r \le r_{\rm c} \\
& B_{0}  (r_{\rm c}/r)^2 &~~~~~& r > r_{\rm c}
\end{aligned}
\end{array}
\right..
\end{equation}
Note that our field model drops off more quickly outside of $r_{\rm c}$ than that of \citet{1999MNRAS.308.1069K} in order that the field will be greatly reduced in magnitude as it approaches the computational grid boundaries.
Additionally, we point out that the Komissarov-type configuration possesses a formally discontinuous magnetic field derivative at $r=r_c$, a feature that we avoid in our other field models (for a discussion of surface currents in initial conditions, see \citealt{2011MNRAS.tmp.1896G}). 
Finally, the force-free case is similar to that used in, \eg \citet{2000A&A...355..818A} and \citet{2009ApJ...700..684M}, featuring a combination of toroidal and poloidal components:
\begin{equation}
B_{\phi}(r)=B_0 \frac{rr_{\rm c}}{(r^2+r_c^2)} ; \hspace{5mm} B_{z}(r)=-B_0 \frac{r_{\rm c}^2}{(r^2+r_c^2)}
\end{equation}
This field configuration corresponds to a constant magnetic pitch of $P=-r_c$.
In the FF models, an envelope function is also applied in the outer $10\%$ of the computational box to ensure that the magnetic field smoothly asymptotes to zero before intersecting the grid boundaries in directions perpendicular to the column axis.
It is worth explicitly noting here that, despite the fact that we provide these expressions in cylindrical coordinates for simplicity, the simulations themselves are conducted on a Cartesian grid.

In all cases, the columns are initialized with a spatially uniform density profile.
The initial pressure profiles for the FF models are also spatially uniform, chosen such that the sound speed is $\sim 0.14$c (except in the case of model FF-hot, in which the sound speed is $\sim 0.57$c).
The PB and RPB models, however, feature pressure profiles that vary radially.
Specifically, the pressure in these two types of simulation is derived from the condition of initial force balance (Eq. \ref{eq:force}), given assumed magnetic field and velocity profiles.
Despite the fact that many of the simulations themselves incorporate special relativistic physics, Equation \ref{eq:force} is sufficiently accurate for initial force balance, which we tested by running the various initial configurations without applying a perturbation to make certain that the equilibrium did not evolve significantly.
Additionally, the pressure is scaled so that the minimum ratio of gas-to-magnetic pressure ($\beta \equiv 2 p/B^2$) achieves the value specified in Table 1.
All models except PB-$\beta$3 and PB-$\beta$30 feature $\beta_{\min} < 1$, reflecting the fact that portions of the computational grid are magnetically dominated.
Additionally, the maximum Alfv\'en speeds for the fiducial models range from $v_{\rm A} \sim 0.1$ (PB) to $v_{\rm A} \sim 0.3$ (FF), the latter of which begins to enter into the relativistically strong field regime.

Finally, the velocity profiles that we employ consist of a combination of large-scale rotational flows ($v_\phi$) and small applied perturbations ($\delta v_{r}, \delta v_{\phi}$).
In all but the RBP models, there is no large-scale rotation, so $v_{\phi}=0$.
In the RBP models, the rotation profile is of the form
\begin{equation}
v_{\phi}(r)=0.5 v_0 (1-\cos[2\pi r/r_{\rm v}]),
\end{equation}
which is identical in form to the sinusoidal magnetic field profiles.
This profile is motivated primarily by desiring rotation that is both slowly increasing from zero near the center of the column (so that it is numerically resolved) and zero also near computational grid boundaries perpendicular to the column axis.
In the fiducial model RPB, $v_0=0.2$c and $r_v=r_c$, so that the magnetic field and rotation field are spatially co-located.
In model RPB-slow, the velocity is reduced such that $v_0=0.1$c (for $r_v=r_c$).
Finally, in model RPB-xrot, the rotation profile is extended radially such that  $r_v=1.5r_c$ (with $v_0=0.2$c), so that there is some rotation exterior to the magnetized region of the column.
We should note that the RPB models are run exclusively using Newtonian physics.
This is primarily for ease of establishing the initial force balance, although we intend to explore the potential effects of special relativistic large-scale rotation and alternative rotation profiles (see, e.g. \citealt{2008A&A...492..621M}) in future work.

The perturbation applied is an $|m|=1$ type perturbation designed to excite the fastest-growing CDI modes:
\begin{equation}
\begin{aligned}
\delta v_r (r,\phi,z) & = \delta v_0 \exp(-r/4) [\sin(\eta)\cos(\phi) + \cos(\eta)\sin(\phi)] \\
\delta v_{\phi} (r,\phi,z) & = \delta v_0 \exp(-r/4) [\cos(\eta)\cos(\phi) -\sin(\eta)\sin(\phi)],
\end{aligned}
\end{equation}
where $\eta \equiv 2\pi z/z_{\rm size}$ with $z_{\rm size}$ representing the size of the domain in the direction parallel to the column axis.
The value of $\delta v_0$ is chosen such that the perturbation magnitude is of order $1\%$ of the initial maximum Alfv\'en speed $v_{\rm A}$ in all but models FF-v6 and FF-v25 (see Table 1).

All of our models are simulated on Cartesian meshes of uniform cubic zone size at all locations on the grid. The computational domain size for the FF models is $x_{\rm size}$/$r_c = y_{\rm size}$/$r_c = 2 z_{\rm size}$/$r_c =32$, where the column axis is initially oriented along the $z$ direction.
In the PB/RPB models, however, we increase the resolution to ensure that the pressure/rotation profiles are adequately resolved.
This has the effect of reducing the domain size for a fixed number of zones, so that for the PB/RPB models the computational grid spans $x_{\rm size}$/$r_c = y_{\rm size}$/$r_c = 2 z_{\rm size}$/$r_c =6.4$.
These domains are spanned by $256\times256\times128$ zones in all of the production-scale models listed in Table 1, and we discuss tests of this resolution in \S\ref{sec:spectral}.
In all models, standard {\tt Athena} outflow boundaries are employed in the $x$ and $y$ directions while periodic boundaries are employed in $z$.
In practice, the simulations are halted before there are substantial interactions with grid boundaries orthogonal to the column axis.

Times in the remainder of this work will be given in units of a representative Alfv\'en crossing time $\tau_{\rm A}$, which we define as $\tau_{\rm A} \equiv r_c/v_{\rm A}$, where $v_{\rm A}$ is the maximum initial Alfv\'en speed on the grid and $r_c$ is the initial column radius.
While the local Alfv\'en speed will obviously be a function of both position and time, we emphasize that our definition of $\tau_{\rm A}$ is only directly indicative of the minimum Alfv\'en crossing time at t=0.

\subsection{Spectral Accuracy \& Numerical Convergence}
\label{sec:spectral}

\begin{figure}
\begin{center}
\leavevmode
\includegraphics[type=pdf,ext=.pdf,read=.pdf,width=\columnwidth]{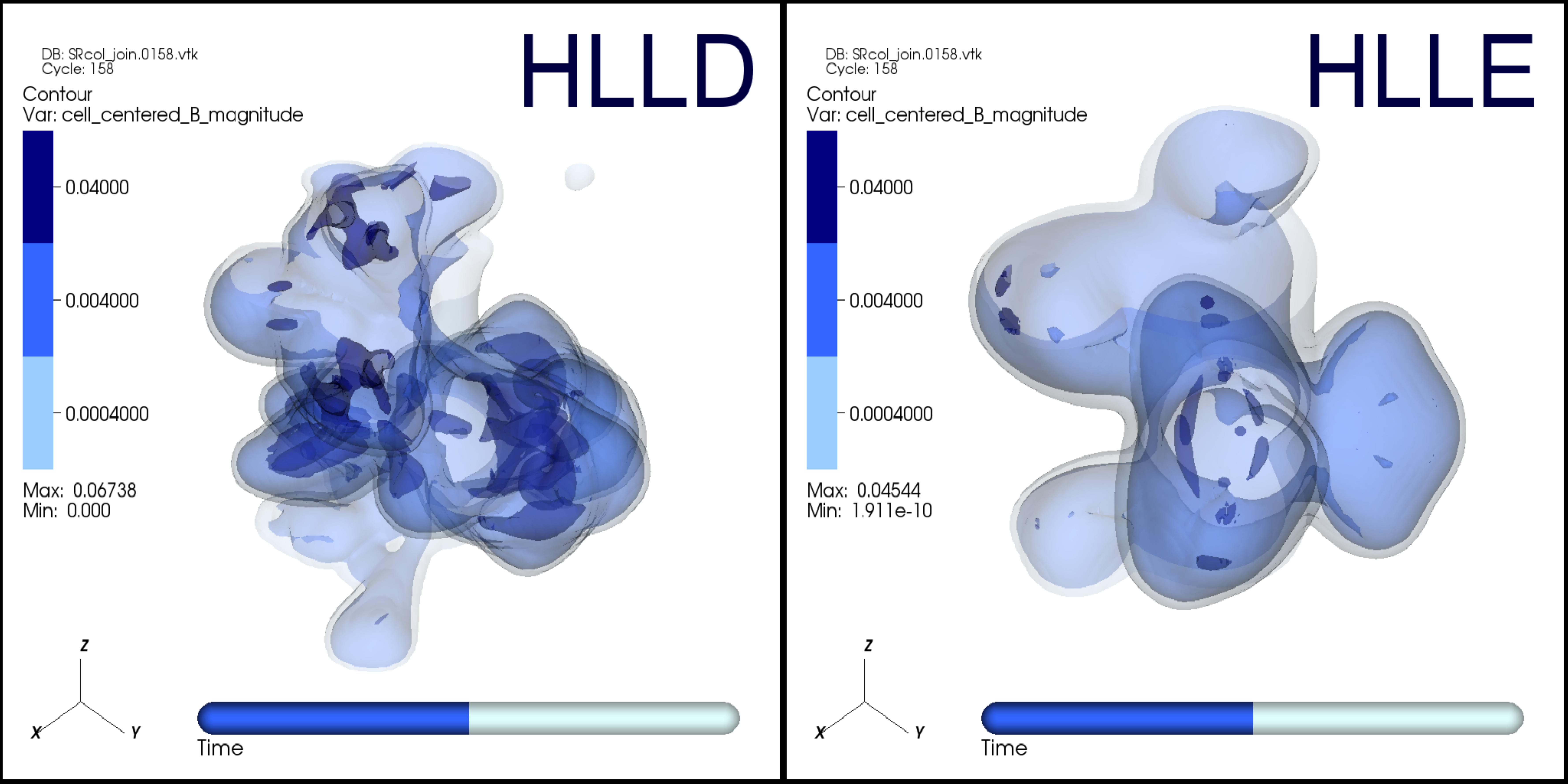}
\end{center}
\caption{A comparison of magnetic field isosurfaces between the HLLD and HLLE solvers for otherwise identical CDI simulations at $t=10~\tau_{\rm A}$.  The resolution used in these models is a factor of two lower than any of our production runs to emphasize how different Riemann solvers can produce strongly divergent outcomes.}
\label{fig:LOWRES_EV}
\end{figure}

As described in \cite{2011ApJS..193....6B} and outlined above, choices regarding the Riemann solver and the order of accuracy of spatial reconstruction (along with numerical resolution) can have important consequences for the interpretation of non-linear MHD phenomena observed in numerical simulations. 
In the case of CDI in relativistic jets, the majority of prior simulations \cite[e.g.][]{2009ApJ...700..684M,2009MNRAS.394L.126M} have been conducted using approximate HLLE-like solvers, where the structure of the Riemann fan is diffused \citep{1983JCoPh..49..357H}. 
For these solvers, discontinuities that are exact solutions of the Riemann problem will evolve via numerical diffusion, even when the flow is at rest \citep{2009MNRAS.393.1141M}. 
This diffusion plays an important role in determining the overall spectral accuracy of the numerical scheme \citep{2009MNRAS.393.1141M,2011ApJS..193....6B}. 
This \emph{numerical} diffusivity can be reduced (and the spectral accuracy of the scheme improved) by including the physics of contact and Alfv\'en waves in the solution of the Riemann problem, e.g. as in the five-wave approximate Riemann solver (which we term `HLLD') initially described by \citet{2005JCoPh.208..315M} in the Newtonian regime and later extended to special relativity by \cite{2009MNRAS.393.1141M}.
If the physics of the problem under consideration depends on the propagation of either of these wave families (e.g. the contact wave in the case of counter-propagating shear flows, or Alfv\'en waves in the case of the magnetorotational instability), then use of Riemann solvers where these waves are diffused is likely to result in incomplete \emph{physical} interpretations of simulation outcomes, even though the solution itself is \emph{mathematically} correct. 
The growth of CDI is driven by the propagation of (modified) Alfv\'en waves, so CDI-driven outcomes are likely affected by choices regarding the Riemann solver. 
This point is illustrated in Figure \ref{fig:LOWRES_EV}, which shows a side-by-side comparison of low-resolution ($128\times128\times64$ zones) test versions of model PB as conducted with the HLLD and HLLE solvers using {\tt Athena} computed using second-order spatial reconstruction (see below). 
Clearly, the CDI-driven flow morphology at this resolution is dramatically affected by the choice of solver, particularly in regions of high magnetization that are conspicuously absent for the model computed with the HLLE solver, a result consistent with those of \cite{2011ApJS..193....6B}.

\begin{figure}
\begin{center}
\leavevmode
\includegraphics[type=pdf,ext=.pdf,read=.pdf,width=\columnwidth]{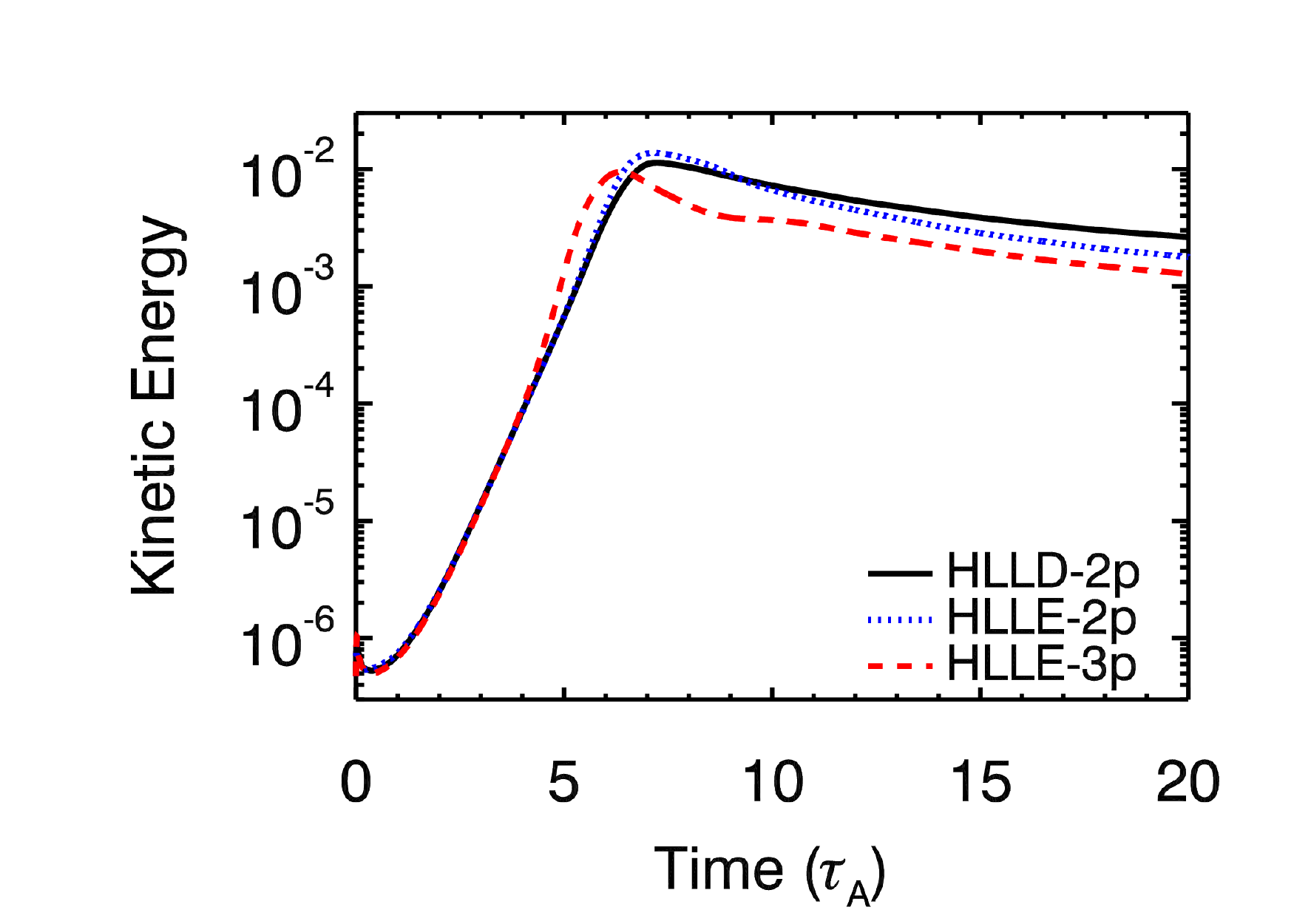}
\end{center}
\caption{Kinetic energy evolution for various solvers and reconstruction orders for the fiducial PB model.  Each model is independently normalized so that the total initial energy (TE+BE+KE) is unity.  The combination of the HLLE solver and 3rd-order (3p) reconstruction leads to a final kinetic energy that is approximately half that of the HLLD solver with 2nd-order (2p) reconstruction.  This is a clear illustration that algorithmic differences can produce divergent behaviors that, in most cases, have stronger effects than even the variation of physical model parameters.}
\label{fig:PB_COMPARE_SOLVER_ENERGY_K}
\end{figure}

\begin{figure}
\begin{center}
\leavevmode
\includegraphics[type=pdf,ext=.pdf,read=.pdf,width=\columnwidth]{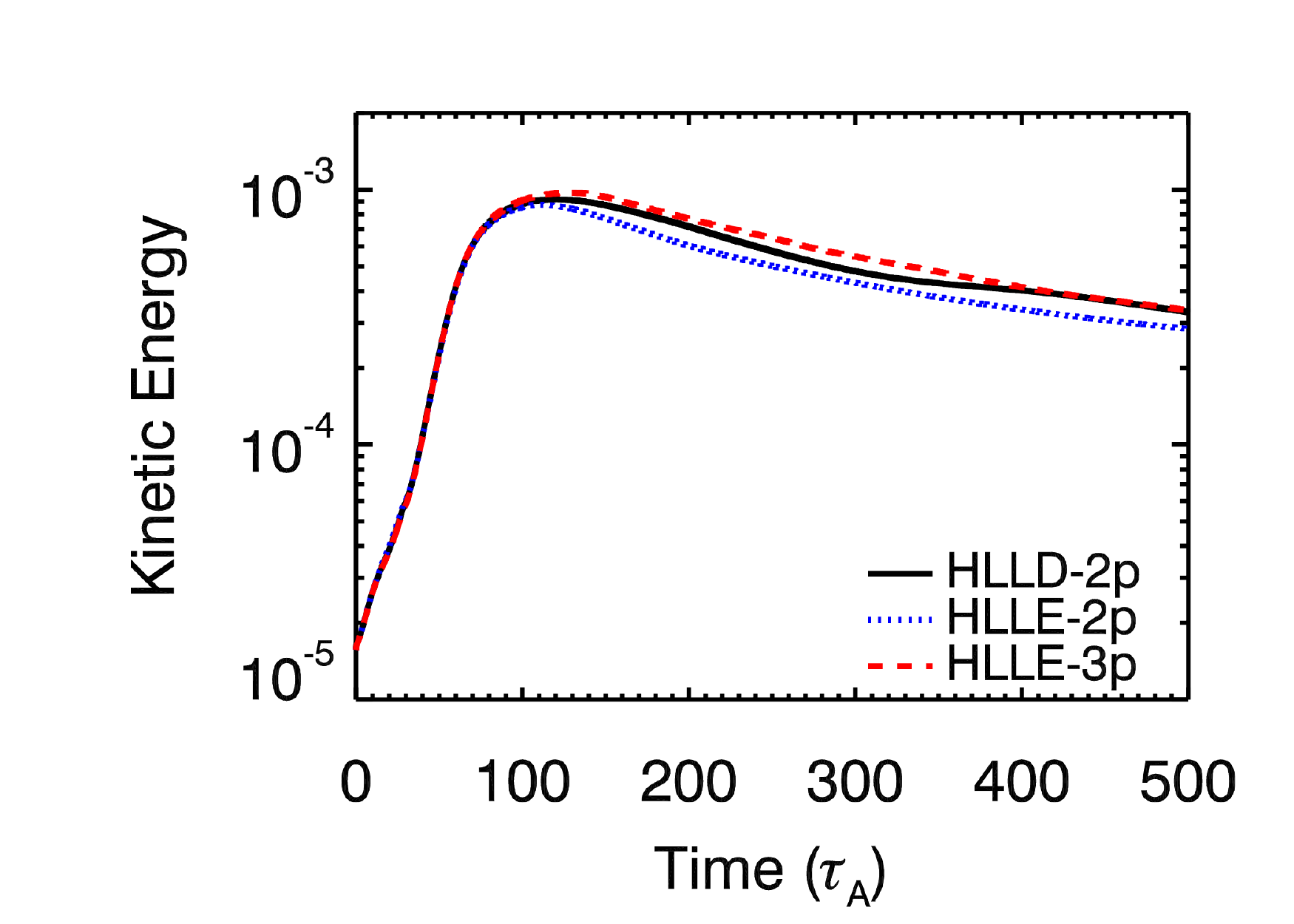}
\end{center}
\caption{Kinetic energy evolution for various solvers and reconstruction orders for the fiducial FF model.  Each model is independently normalized so that the total initial energy (TE+BE+KE) is unity.  The final energies differ by at most $\sim17\%$, which is significant but much less pronounced than in the PB model.}
\label{fig:FF_COMPARE_SOLVER_ENERGY_K}
\end{figure}

To illustrate that this is not a problem exclusively confined to low resolutions, we also present a comparison of energy evolution for the full-resolution PB models described in Section \ref{sec:initialconditions} and listed in Table 1.
While we will defer an extensive discussion of CDI energetics to Section \ref{sec:results}, it will suffice here to note that instability manifests itself as an initially exponential growth of kinetic energy (the so-called ``linear'' growth phase) that occurs at the expense of magnetic energy until the instability achieves a non-linear saturated state.
Figures \ref{fig:PB_COMPARE_SOLVER_ENERGY_K} and \ref{fig:FF_COMPARE_SOLVER_ENERGY_K} illustrate sample dependences of the kinetic energy evolution on solver.
In Figure \ref{fig:PB_COMPARE_SOLVER_ENERGY_K}, we see that solutions computed with the HLLD and HLLE solvers (again using 2nd-order spatial reconstruction) exhibit nearly identical growth rates during the linear phase of CDI development in model PB.
Near the peak of the linear CDI development, however, the two solutions begin to diverge slightly until the end of the simulations, at which point the final kinetic energy for case computed with the  HLLD solver is approximately 1.5 times that of the HLLE case.
In the FF models, the difference between solutions computed with the two solvers is significantly smaller, only $\sim~17\%$ over a much larger number of crossing times. 
Again, these differences are entirely attributable to the choice of Riemann solver for configurations that were otherwise identical. 
To summarize, improving the spectral accuracy of the Riemann solver leads to dramatic changes in flow morphology and energetics; in particular the efficiency at which magnetic energy is converted into kinetic energy. 
Since the HLLD approximate Riemann solver contains a more physically complete description of the breakdown of the Riemann problem at the cell interface for the RMHD equations than the HLLE approximate Riemann solver, we adopt the HLLD solver as our standard of comparison and utilize it for all of the simulations presented in \S\ref{sec:results}.


While the algorithms implemented in {\tt Athena} are overall second-order accurate in space and time, extensive testing has revealed that use of third-order spatial reconstruction can lower truncation error and increase the accuracy of the solution \citep{2008ApJS..178..137S}. 
One might expect therefore, that the divergence in flow morphology and energy evolution between solutions computed using the HLLE and HLLD approximate Riemann solvers can be ameliorated by improving the accuracy of spatial reconstruction from second to third order for the HLLE Riemann solver. 
The data of Figures \ref{fig:PB_COMPARE_SOLVER_ENERGY_K} and \ref{fig:FF_COMPARE_SOLVER_ENERGY_K} show that, in the PB models, the asymptotic value of the kinetic energy using 2nd-order reconstruction is a factor of $\sim 1.4$ times that of 3rd-order reconstruction, while for the FF models, the ordering is reversed, with 3rd-order reconstruction producing kinetic energies roughly $18\%$ greater than 2nd-order. 
From this, we conclude that while the use of third-order reconstruction in combination with the HLLE solver does reduce differences in flow energetics between Riemann solvers in the FF case, the more complex dynamics introduced when (e.g.) gas pressure plays a significant role in force balance are best captured by the use of the more physically complete HLLD approximate Riemann solver, even at lower spatial accuracy. 
This result is consistent with expectations derived from extensive comparisons between high-order (e.g. WENO5) schemes and second-order Godunov methods, where it was demonstrated that, for non-linear problems, both high order schemes and second-order Godunov methods converge at first order \citep{2004JCoPh.196..259G}. 
These authors conclude that, while higher order methods deliver higher accuracy solutions at fixed computational cost for smooth solutions, for non-linear problems second-order Godunov methods deliver higher accuracy solutions than higher order schemes (again, at fixed computational cost).
As a result, we adopt second-order spatial reconstruction for the all of the simulations presented in \S\ref{sec:results}.

\begin{figure}
\begin{center}
\leavevmode
\includegraphics[type=pdf,ext=.pdf,read=.pdf,width=\columnwidth]{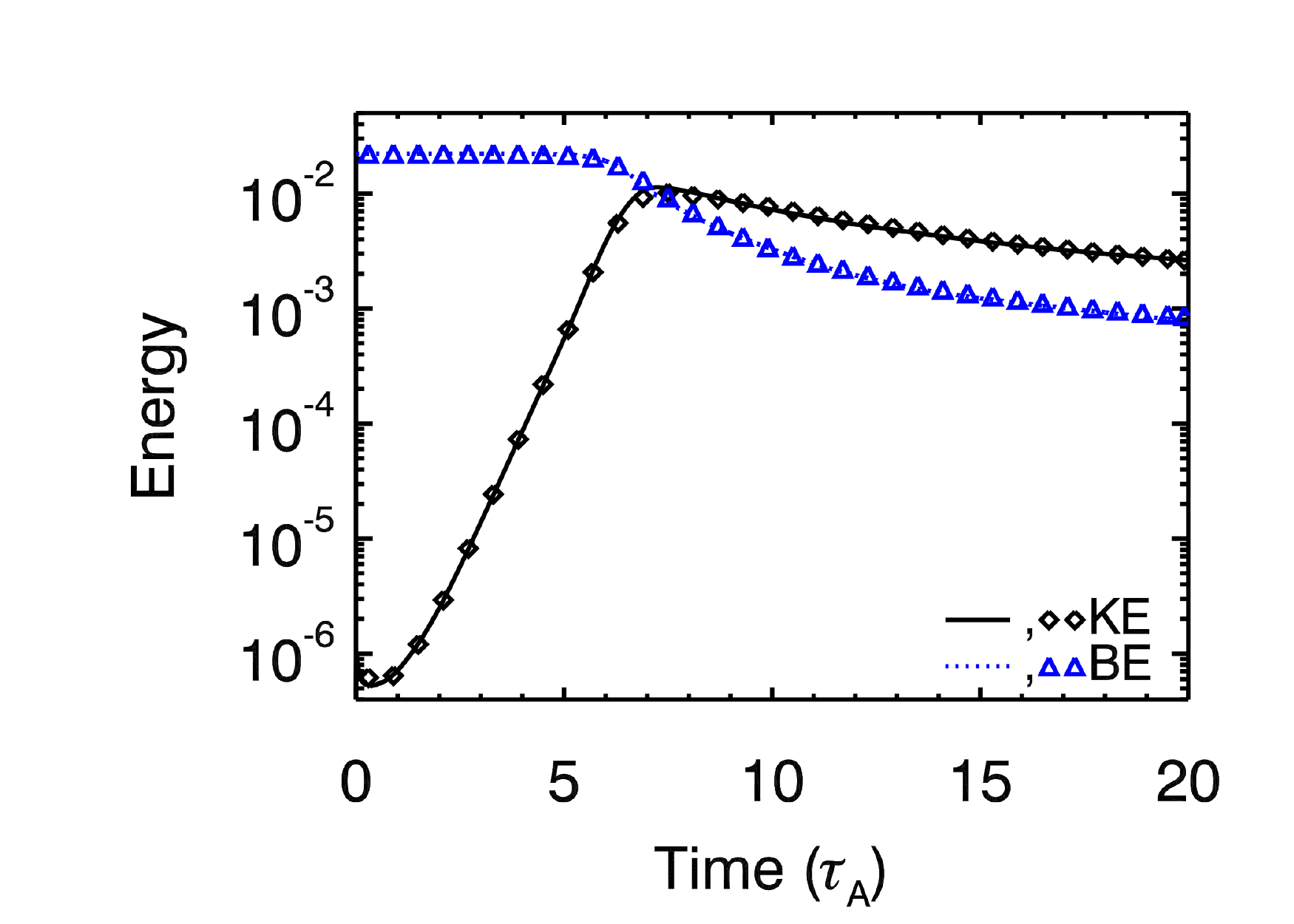}
\end{center}
\caption{A resolution study of energetics for the PB models, comparing the fiducial model PB (lines) with a model in which the physical resolution is doubled (glyphs).  Each model is independently normalized so that the total initial energy (TE+BE+KE) is unity.  The models exhibit good convergence, suggesting that the evolution of small-scale structure does not dominate the energetics of the system.}
\label{fig:PB_COMPARE_HIRES_ENERGY_KB}
\end{figure}

\begin{figure}
\begin{center}
\leavevmode
\includegraphics[type=pdf,ext=.pdf,read=.pdf,width=\columnwidth]{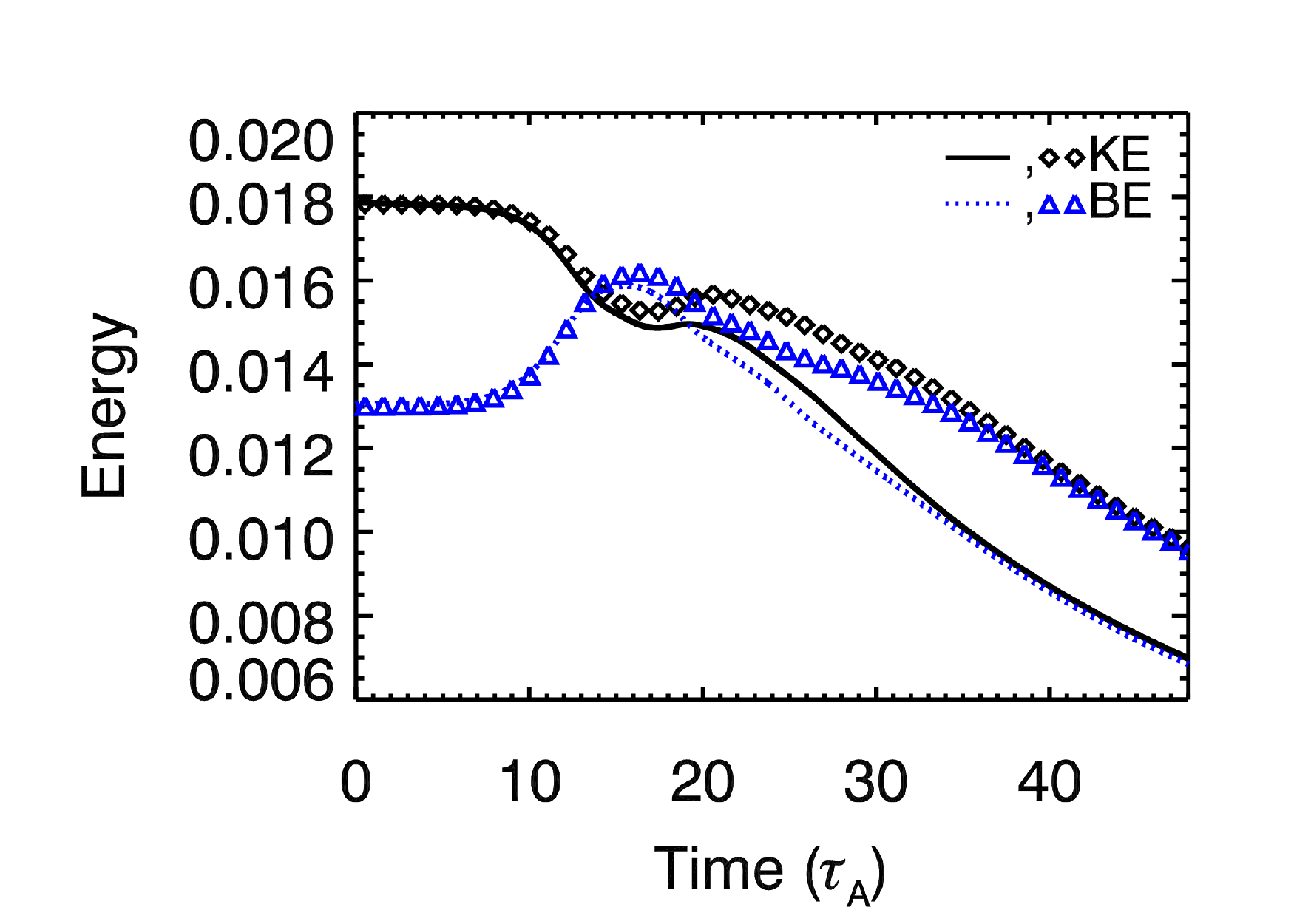}
\end{center}
\caption{A resolution study of energetics for the RPB models, comparing the fiducial model RPB (lines) with a model in which the physical resolution is doubled (glyphs).  Each model is independently normalized so that the total initial energy (TE+BE+KE) is unity.  The models exhibit adequate convergence with maximum deviations of approximately $5\%$ near the onset of saturation and asymptotic differences in the non-linear regime of approximately $30-35\%$.}
\label{fig:RPB_COMPARE_HIRES_ENERGY_KB}
\end{figure}

\begin{figure}
\begin{center}
\leavevmode
\includegraphics[type=pdf,ext=.pdf,read=.pdf,width=\columnwidth]{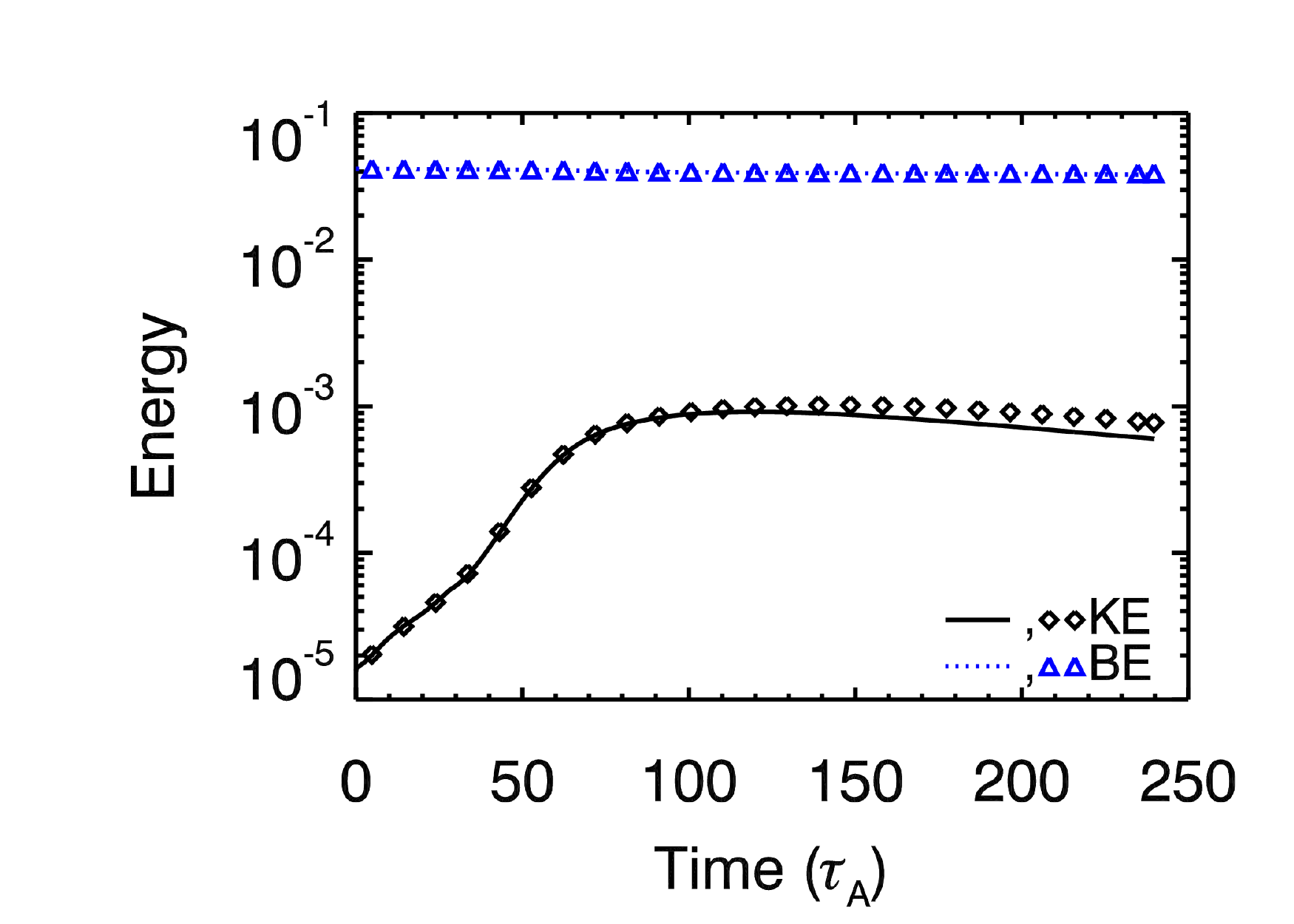}
\end{center}
\caption{A resolution study of energetics for the FF models, comparing the fiducial model FF (lines) with a model in which the physical resolution is doubled (glyphs).  Each model is independently normalized so that the total initial energy (TE+BE+KE) is unity.  The models exhibit good convergence during the linear growth phase, but the kinetic energy evolves differently at the $\sim 30\%$ level well after CDI saturation.}
\label{fig:FF_COMPARE_HIRES_ENERGY_KB}
\end{figure}

\begin{figure*}
\begin{center}
\leavevmode
\includegraphics[type=pdf,ext=.pdf,read=.pdf,width=\textwidth]{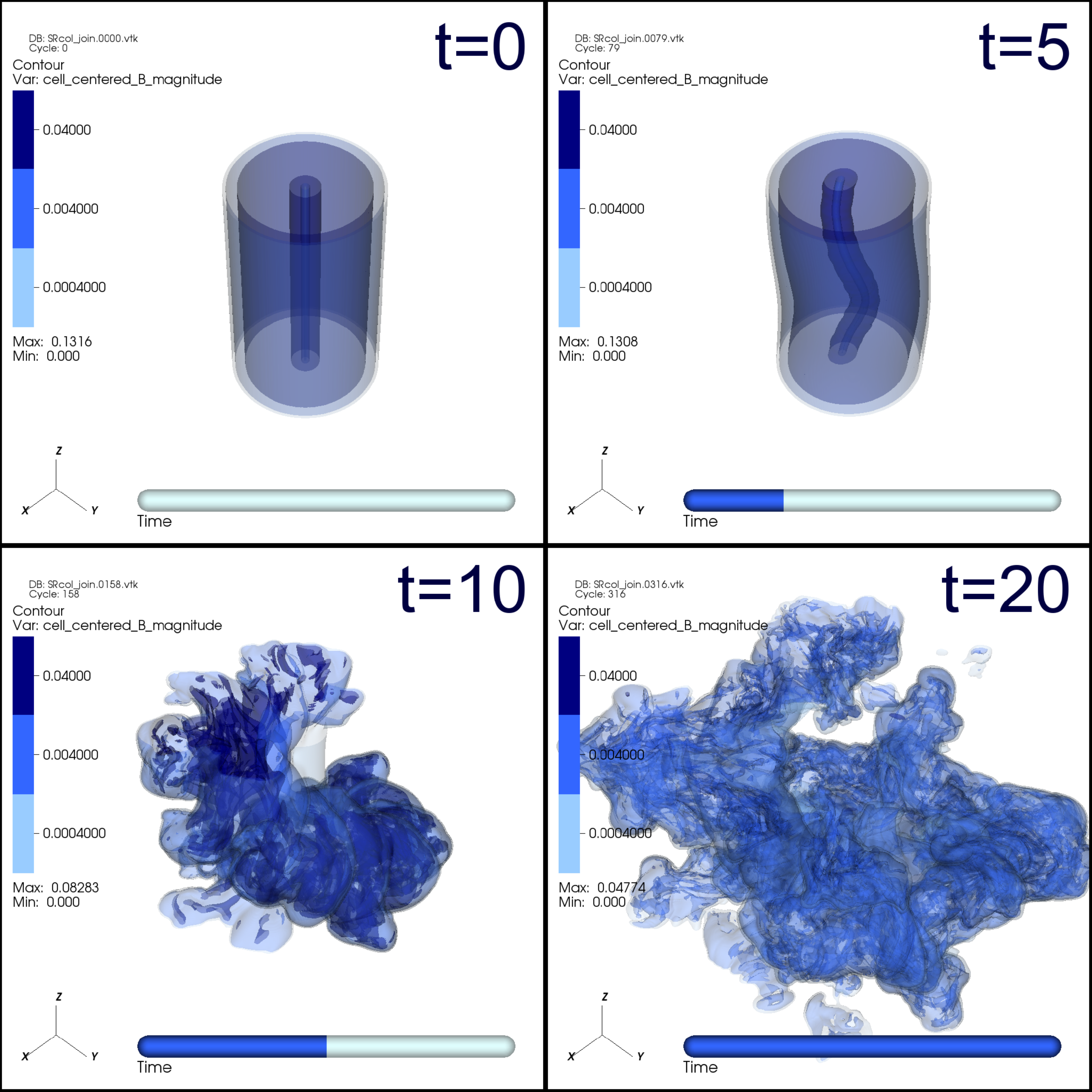}
\end{center}
\caption{Magnetic field isosurfaces for the fiducial PB model at four different times (specified in units of $\tau_{\rm A}$).  All variants of the PB model are eventually subject to CDI, becoming completely disrupted over time.}
\label{fig:PB_EV}
\end{figure*}

\begin{figure*}
\begin{center}
\leavevmode
\includegraphics[type=pdf,ext=.pdf,read=.pdf,width=\textwidth]{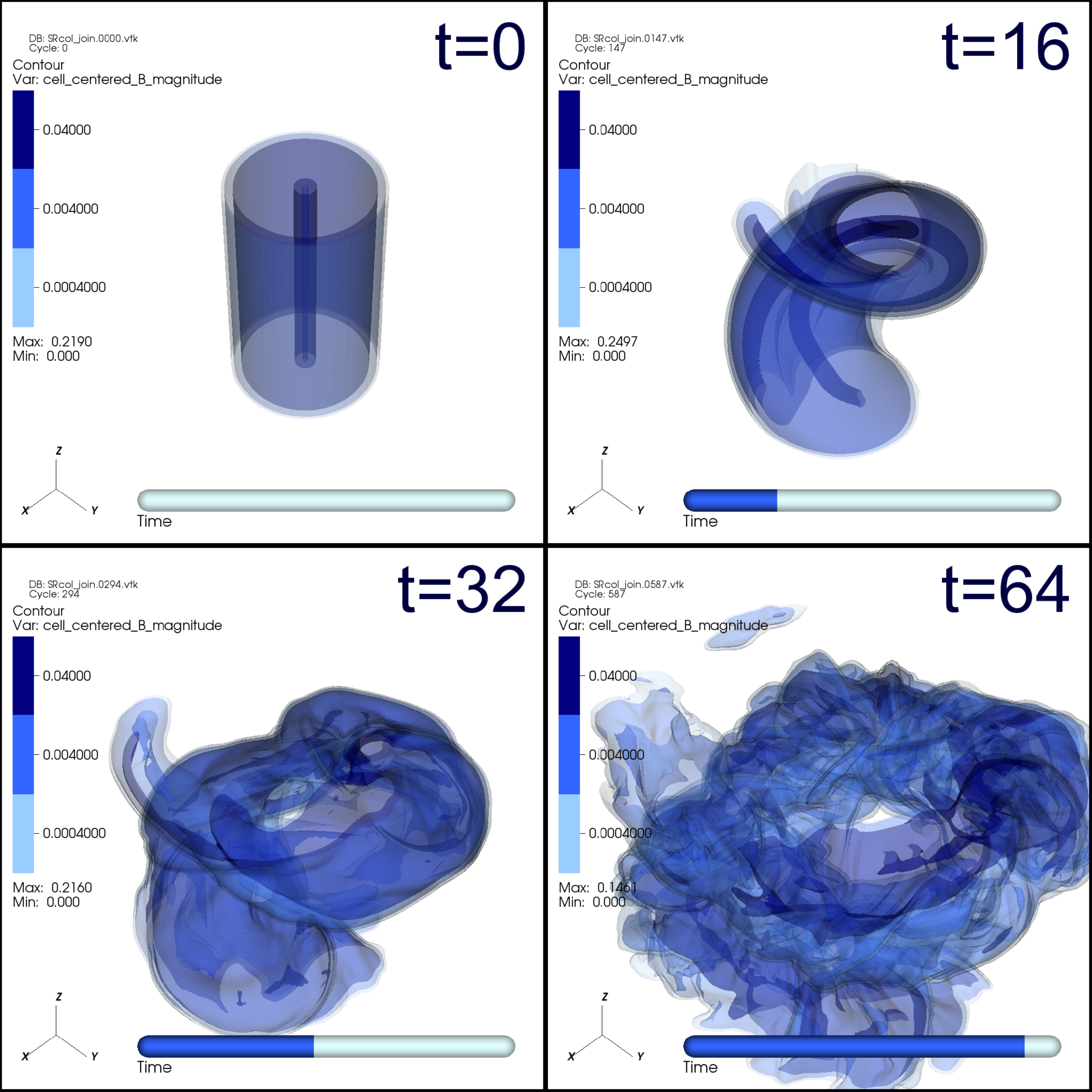}
\end{center}
\caption{Magnetic field isosurfaces for the fiducial rotating RPB model at four different times (specified in units of $\tau_{\rm A}$).  As is the case for the PB models, the RPB models eventually become disrupted, although their field strengths remain generally higher than those of the PB models.}
\label{fig:RPB_EV}
\end{figure*}

The simulations presented here were conducted in ideal relativistic MHD, and, as a result, none of the solutions in the nonlinear regime can be regarded as `converged'. 
For convergence, a physical dissipation scale (provided by either, e.g., a Navier-Stokes viscosity or Ohmic resistivity) would have to be included in the problem \citep{2011ApJS..193....6B}. 
Such calculations are extremely challenging in relativistic MHD \cite[see e.g.][]{2011ApJ...735..113T} and are well beyond the scope of the current work. 
In addition, it is likely that in relativistic, magnetized astrophysical jets, the dissipation scale would be separated by many orders of magnitude from the smallest scales accessible by current computational resources. 
Instead, convergence of the solution must be judged during the linear growth phase of CDI. 
Using the choices of Riemann solver and order of spatial reconstruction outlined previously (i.e. the HLLD approximate Riemann solver and second order spatial reconstruction), we compute simulations of the PB, RPB and FF cases described in \S\ref{sec:initialconditions} at both the fiducial resolution ($256\times256\times128$ zones) and at twice this resolution (using otherwise identical configurations). 
The data of Figure \ref{fig:PB_COMPARE_HIRES_ENERGY_KB}--\ref{fig:FF_COMPARE_HIRES_ENERGY_KB} demonstrate that the linear growth phase associated with each class of initial condition (PB: $t\lesssim7\tau_A$; RPB: $t\lesssim15\tau_A$; FF: $t\lesssim75\tau_A$) is well converged at the fiducial resolution of $256\times256\times128$ zones, which we utilize for the remainder of this work. 
While the subsequent, non-linear, stage of the evolution cannot be used to judge convergence of the models, it is instructive to examine the differences that arise at higher resolution. 
Model PB exhibits non-linear evolution that differs by $\lesssim5\%$ between the fiducial and higher resolutions; both quantitative and qualitative conclusions drawn from the fiducial resolution during the non-linear evolution will therefore remain unchanged at higher resolution for this model. 
Models RPB and FF exhibit non-linear evolutions that differ by $\sim35\%$ between the two resolutions at late times.
In both cases, the qualitative non-linear evolution is not strongly affected by resolution, however. 
For the RPB model in particular, the ratio of the magnetic-to-kinetic energy remains qualitatively similar (i.e. approximate equipartition) during the non-linear stage. 
As a result, we conclude that while quantitative conclusions drawn concerning the non-linear behaviors of models RPB and FF will be subject to resolution effects, qualitative conclusions regarding the physics at work in these models will be resolution \emph{independent}.

\section{Results}
\label{sec:results}
\subsection{Morphology}
\begin{figure*}
\begin{center}
\leavevmode
\includegraphics[type=pdf,ext=.pdf,read=.pdf,width=\textwidth]{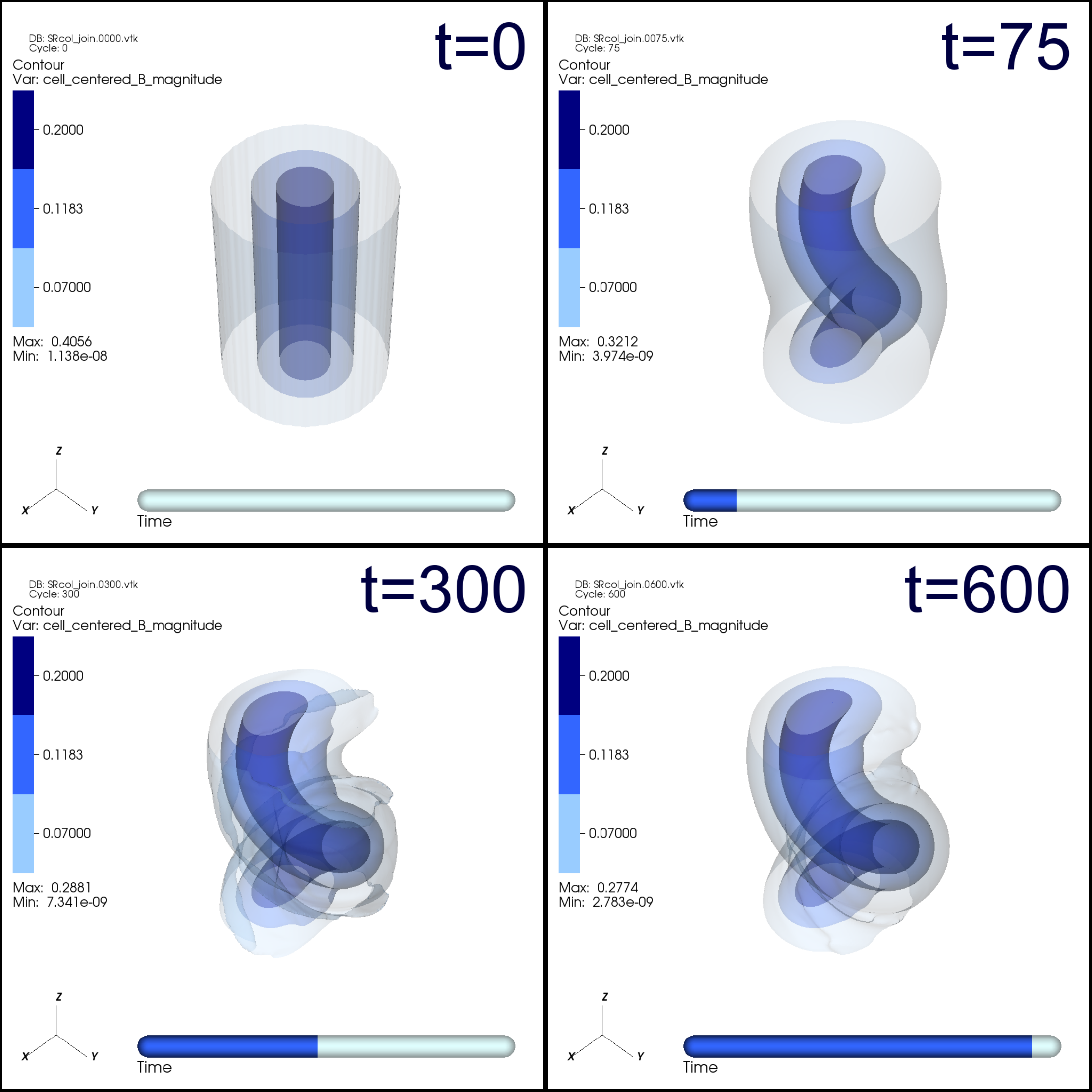}
\end{center}
\caption{Magnetic field isosurfaces for the fiducial FF model at four different times (specified in units of $\tau_{\rm A}$).  Unlike the PB/RPB models, the FF models are merely deformed, rather than completely disrupted.  Although the system continues to evolve slowly after $t\sim 100~\tau_{\rm A}$, the gross morphology does not change dramatically after this time.}
\label{fig:FF_EV}
\end{figure*}

To begin, we will discuss the distinct morphological evolutions that result from the three different classes of initial force balance.
Here, we will focus primarily on comparing the fiducial models (PB, RPB, and FF), deferring a detailed discussion of the other models to our analysis of the system energetics. 
Figures \ref{fig:PB_EV}-\ref{fig:FF_EV} show snapshots of the evolution of magnetic field strength in models PB, RPB, and FF.
The fields are shown as isosurfaces, which is sufficient to illustrate the degree to which the field structures bend and/or become disrupted as the systems evolve.

Figures \ref{fig:PB_EV} and \ref{fig:RPB_EV} show that both the PB and RPB columns are strongly disrupted, eventually showing clear signs of turbulent evolution.
Model PB, in particular, features field structures that emerge on a large variety of spatial scales, ranging from the size of the system to smaller scales that are obviously not directly relatable to the initial $|m|=1$ scale perturbation.
Additionally, the PB column begins to become disarranged quite rapidly, in less than $10~\tau_{\rm A}$, and achieves some degree of turbulence by $ 20~\tau_{\rm A}$.
Considering that the system has also expanded to become larger than the initial column radius used to characterize $\tau_{\rm A}$, this is rapid evolution indeed.

\begin{figure*}
\begin{center}
\leavevmode
\includegraphics[type=pdf,ext=.pdf,read=.pdf,width=0.33\textwidth]{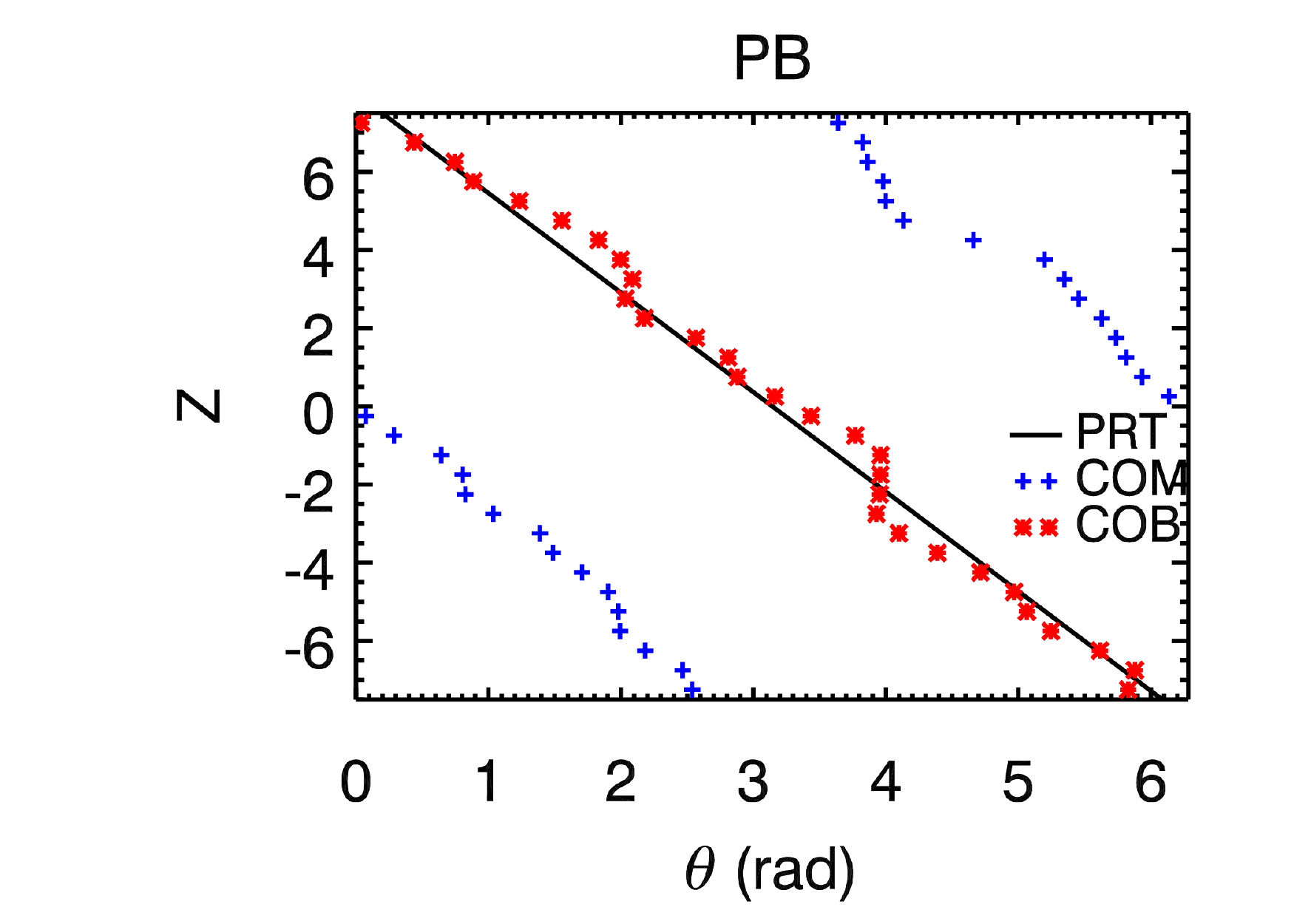}
\includegraphics[type=pdf,ext=.pdf,read=.pdf,width=0.33\textwidth]{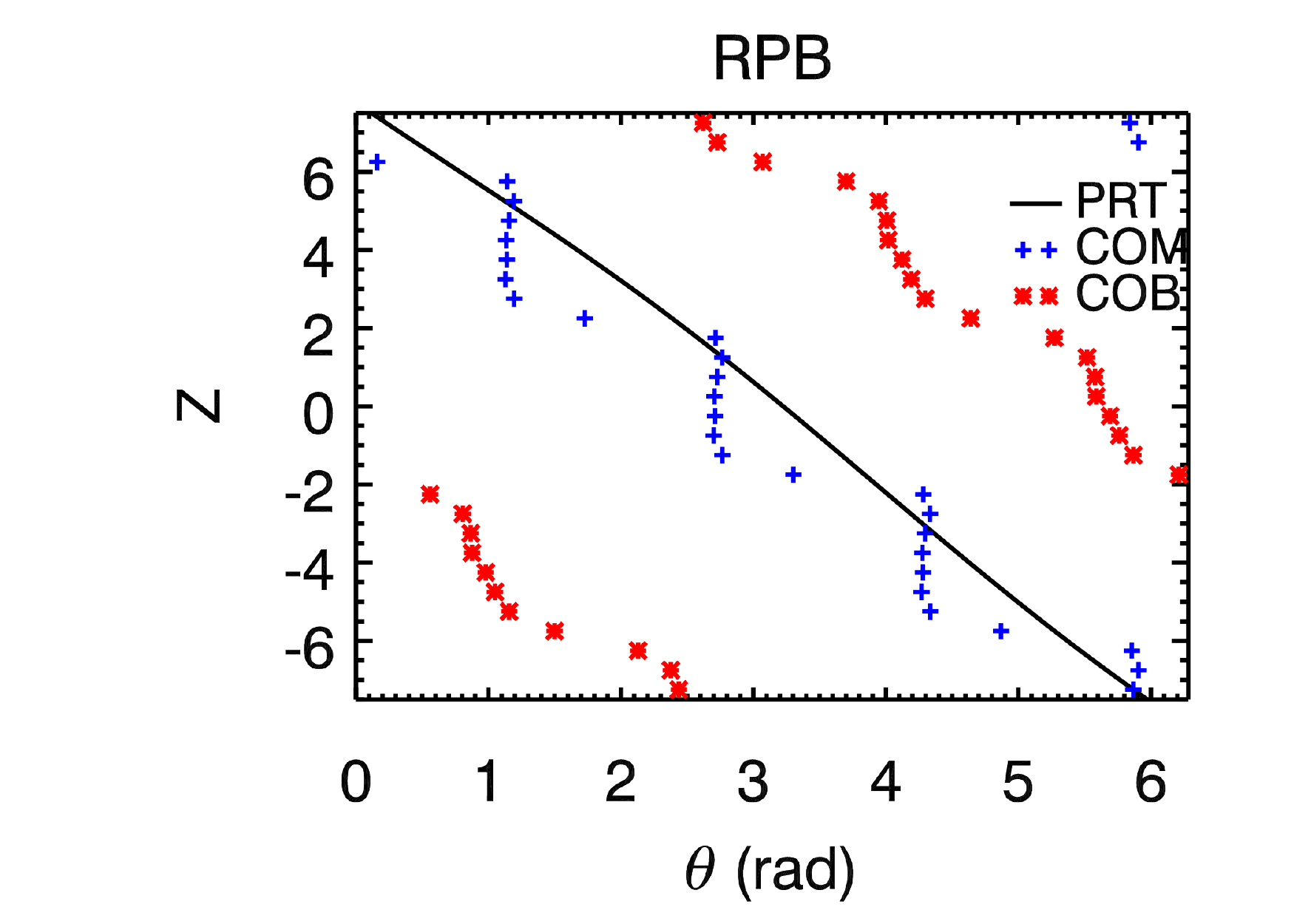}
\includegraphics[type=pdf,ext=.pdf,read=.pdf,width=0.33\textwidth]{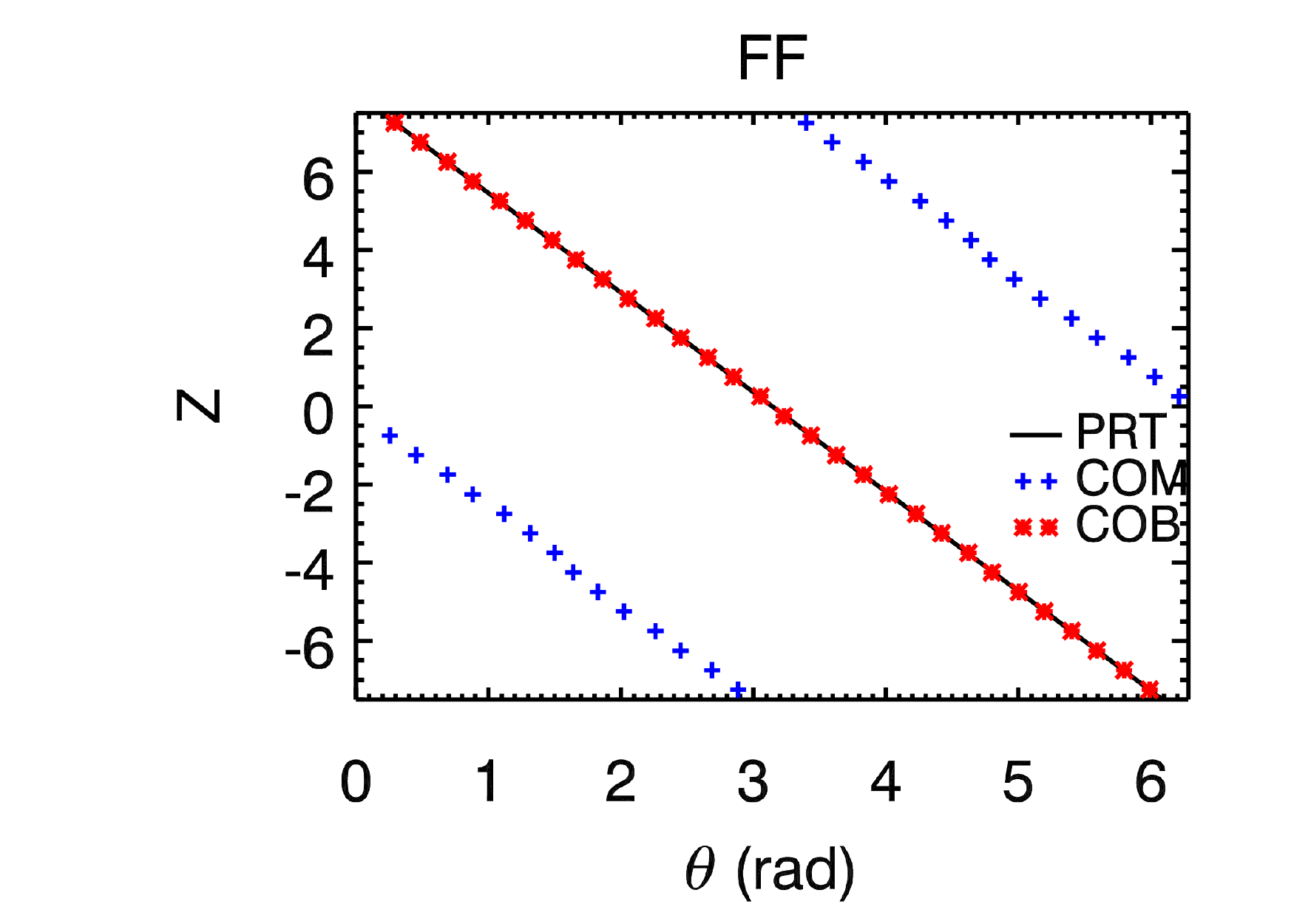}
\end{center}
\caption{Relating the direction of column deformation to the initial perturbation for the PB model (left), the RPB model (center), and the FF model (right).
Solid lines show the angular orientation of the initial perturbation as a function of z, where the glyphs depict the center-of-mass (COM) and the average location of the magnetic field (COB) at late times.
In all cases, the magnetic field and center-of-mass are close to $180$ degrees out of phase, which is not surprising since the field expands into a region of initially uniform density.
For the PB and FF models, the magnetic field is clearly correlated with the direction of the initial perturbation, as we would expect.
The RPB model features an evolved state that is shifted in phase as a result of the system's rotation.}
\label{fig:PERTANGLES}
\end{figure*}

The RPB model, too, becomes disrupted, with notable differences from the PB model.
One dissimilarity is a result of the system's rotation.
Because the perturbation is initially non-axisymmetric, rotation strongly affects the orientation of the initial deformation and subsequent evolution.
We also find that the RPB model appears to deform less rapidly than model PB.
One might naively have assumed that the RPB model would only evolve more rapidly than model PB, given that the two systems have similar initial field configurations and that rotation provides an additional source of free energy in the RPB system.
In detail, however, the two models also feature different initial pressure gradients to satisfy Equation \ref{eq:force}.
This can further affect the magnetic field strength since it is only the minimum value of $\beta$ that is constrained, meaning that this value can be achieved at different locations in different models.
That the RPB and PB models evolve over different timescales simply illustrates that the characteristic Alfv\'en crossing time alone may not always be perfectly indicative of the CDI evolution timescale, particularly if the system is undergoing significant evolution.
We further note that the final field structure in model RPB does not appear to have accessed as large a range of spatial scales as model PB.
This is again likely to be a result of the slower evolution of this system. 

Finally, Figure \ref{fig:FF_EV} illustrates that the FF model is deformed but not completely disrupted.
Specifically, it is clear that the column reacts to the initial perturbation, but it is equally obvious that no smaller scales develop appreciable power as the system evolves.
Furthermore, there is very little morphological change in the system after $t \sim 100-200~\tau_{\rm A}$, especially where the strongest field is located.
This is obviously a completely different regime of behavior from what was seen in those models that achieved initial force balance through a combination of magnetic and pressure forces.
Whereas model PB becomes largely disrupted by $20~\tau_{\rm A}$, model FF continues to be dominated by the $|m|=1$ large-scale deformation even after evolving for $600~\tau_{\rm A}$ and shows no sign of turbulence.

To demonstrate that we are in fact seeing these systems respond to the applied perturbation, we show in Figure \ref{fig:PERTANGLES} the relationship between the deformation of the column and the direction of the applied perturbation as a function of the position along the column axis.
Specifically, we look at the angular position of both the center-of-mass (COM), defined in a given direction as $\Sigma \rho_i x_i/\Sigma \rho_i$, and the analogous quantity for the magnetic pressure (COB), defined as  $\Sigma B^2_i x_i/\Sigma B^2_i$.
Here, the Newtonian expressions for the density, position, and the magnetic field are adequate to provide a qualitative sense of whether or not the perturbation and subsequent evolution are aligned.
For models PB and FF, Figure \ref{fig:PERTANGLES} illustrates that the orientations of the COB (asterisks) and the initial perturbation (lines) are well-correlated at all heights within in the column.
In contrast, the COM (plus signs) is out of phase by $\pi$ radians in these two models.
Together, these facts suggest that the magnetic fields move in the direction of the initial perturbation while the mass moves in the opposite direction.
This is not terribly surprising, given that a deformed set of magnetic field isosurfaces will generally expand to bound a larger volume than initially cylindrical isosurfaces.
The density inside such a region is observed to decrease accordingly as a consequence of this expansion (and the magnetic flux freezing captured in ideal MHD), leading to the center of mass becoming displaced in the opposite direction.
In practice, this process leads to only a small overall displacement of the center of mass because so much of the mass in the outer portions of the grid is located in regions of weak or zero magnetization and subsequently does not move. 
In model RPB, we see similar behavior as in model PB, but shifted out of phase as a result of the system's rotation.
In this model, different time snapshots would naturally feature different phase shifts.

\subsection{Energetics}
To understand how and why these different morphologies evolve, we now examine the details of energy flow in these systems.
Our primary goal is to determine the degree to which the magnetic energy is being converted into kinetic energy as would be indicative of CDI.
Furthermore, we seek to determine whether or not this behavior saturates and what levels of kinetic energy amplification are attained through the action of CDI.
Before proceeding, we should define precisely what we mean when discussing the various forms of energy.
In practice, {\tt Athena} treats the total energy, defined as $\rho h \Gamma^2 -p + |\mathbf{B}|^2/2 + [|\mathbf{v}|^2|\mathbf{B}|^2 - (\mathbf{v \cdot B})^2]/2$, where $h$ is the relativistic enthalpy and $\Gamma$ is the Lorentz factor (\eg \citealt{2007ApJS..170..228M}, \citealt{2011ApJS..193....6B}).
Subtracting off the rest mass energy, the kinetic energy is $KE=(\Gamma-1)\Gamma\rho$.
The magnetic energy is taken to be all terms containing the magnetic field, such that $BE=|\mathbf{B}|^2/2 + [|\mathbf{v}|^2|\mathbf{B}|^2 - (\mathbf{v \cdot B})^2]/2$.
In our models, the velocity terms in the magnetic energy expression are typically measured to be less than one percent of the total magnetic energy, indicative of the characteristically low velocities achieved in these flows.
The thermal energy is the remainder after the rest mass, kinetic energy, and thermal energies have been subtracted from the total energy.

\subsubsection{Fiducial Models}
\begin{figure}
\begin{center}
\leavevmode
\includegraphics[type=pdf,ext=.pdf,read=.pdf,width=\columnwidth]{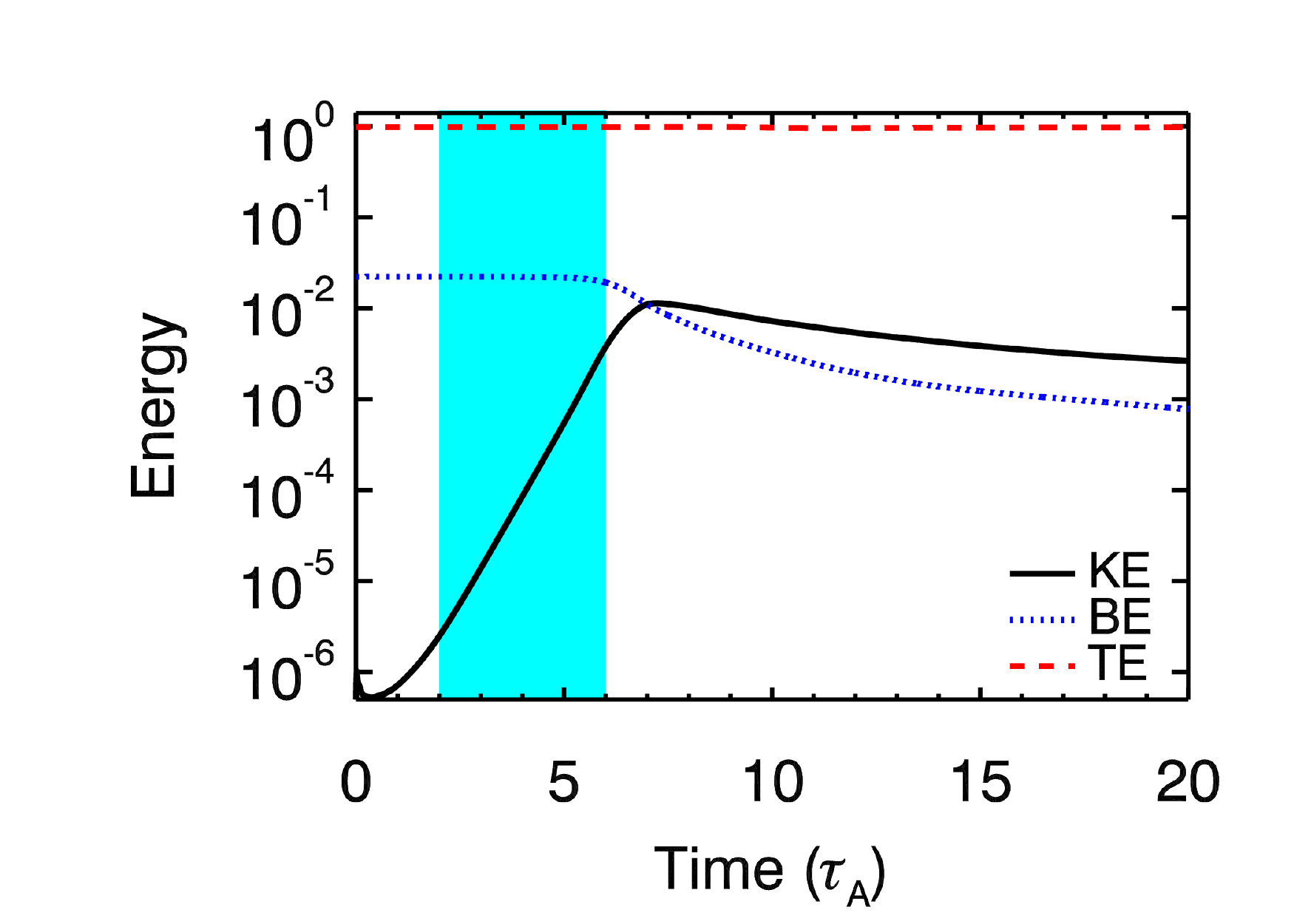}
\end{center}
\caption{Energy evolution over time for the fiducial PB model, normalized so that the total initial energy (TE+BE+KE) is unity, where TE=thermal energy, BE=magnetic energy, and KE=kinetic energy.  The exchange of magnetic into kinetic energy reflects the development of CDI, which appears to saturate after several $\tau_{\rm A}$.  The cyan region indicates the approximate time range over which CDI experience linear growth.}
\label{fig:PB_SINGLE_ENERGY_KBT}
\end{figure}

\begin{figure}
\begin{center}
\leavevmode
\includegraphics[type=pdf,ext=.pdf,read=.pdf,width=\columnwidth]{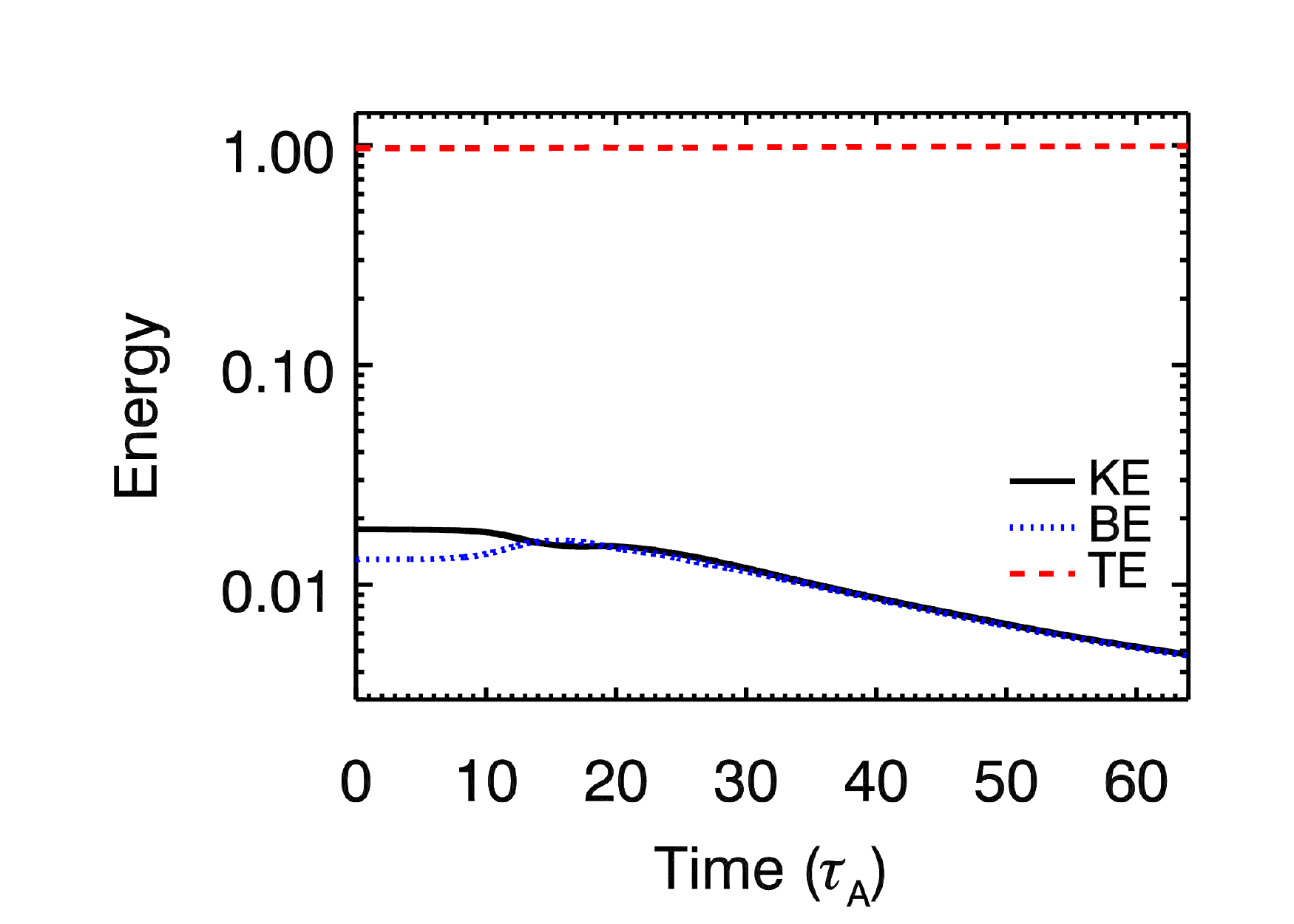}
\end{center}
\caption{Energy evolution over time for the fiducial RPB model, normalized so that the total initial energy (TE+BE+KE) is unity.  In this case, the total kinetic energy (which includes rotational energy) actually decreases and kinetic/magnetic energy follow the same trend after a few tens of $\tau_{\rm A}$, both decreasing as they dissipate into thermal energy.}
\label{fig:RPB_SINGLE_ENERGY_KBT}
\end{figure}

\begin{figure}
\begin{center}
\leavevmode
\includegraphics[type=pdf,ext=.pdf,read=.pdf,width=\columnwidth]{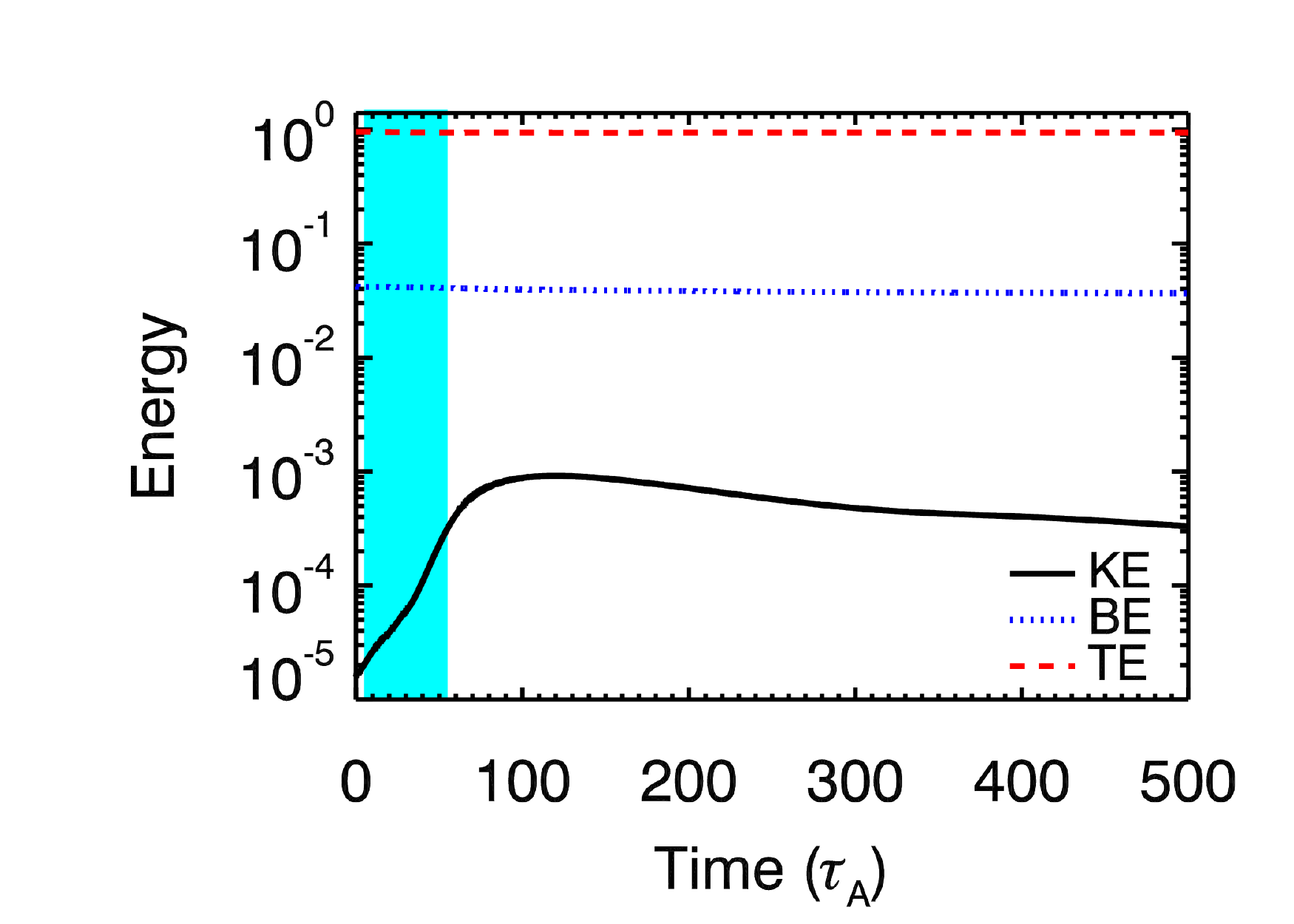}
\end{center}
\caption{Energy evolution over time for the fiducial FF model, normalized so that the total initial energy (TE+BE+KE) is unity.  Although the CDI initially act to amplify the kinetic energy at the expense of magnetic energy, the overall levels of kinetic energy are always far below those of the magnetic and kinetic energy (unlike the fiducial PB and RPB models).   The cyan region indicates the approximate time range over which the CDI experience linear growth.}
\label{fig:FF_SINGLE_ENERGY_KBT}
\end{figure}

Figures \ref{fig:PB_SINGLE_ENERGY_KBT}-\ref{fig:FF_SINGLE_ENERGY_KBT} show the energy evolution for models PB, RPB, and FF.
In each case, the figures depict the average energy contained in each form (kinetic, magnetic, and thermal) as measured over the entire computational grid and normalized such that the total initial energy is unity.
Although some energy exits through the open computational boundaries at late times as these models evolve, this normalization reflects the total energy measured at any given time with an accuracy of better than two percent for all of our fiducial simulations.
For all three fiducial models, the initial energy distribution is dominated by thermal energy with the magnetic energy comprising at most a few percent of the total grid energy.
Of course, one should keep in mind that this relative ordering is largely an artifact of the grid size convolved with the radial profile of initial magnetic energy.
Clearly, magnetic energy can make up a substantially larger fraction of the total energy in local regions where the field is stronger, and in fact all fiducial models feature a minimum $\beta < 1$.
In each model, the kinetic energy attributable to the applied perturbation is a very small fraction of the initial total energy, as can be seen in Figures \ref{fig:PB_SINGLE_ENERGY_KBT} and \ref{fig:FF_SINGLE_ENERGY_KBT}.
The perturbation is equally small for model RPB (Figure \ref{fig:RPB_SINGLE_ENERGY_KBT}), but that model also features a non-negligible (at the $2\%$ level) amount of initial kinetic energy reflecting the rotation profile.

\begin{figure*}
\begin{center}
\leavevmode
\includegraphics[type=pdf,ext=.pdf,read=.pdf,width=0.33\textwidth]{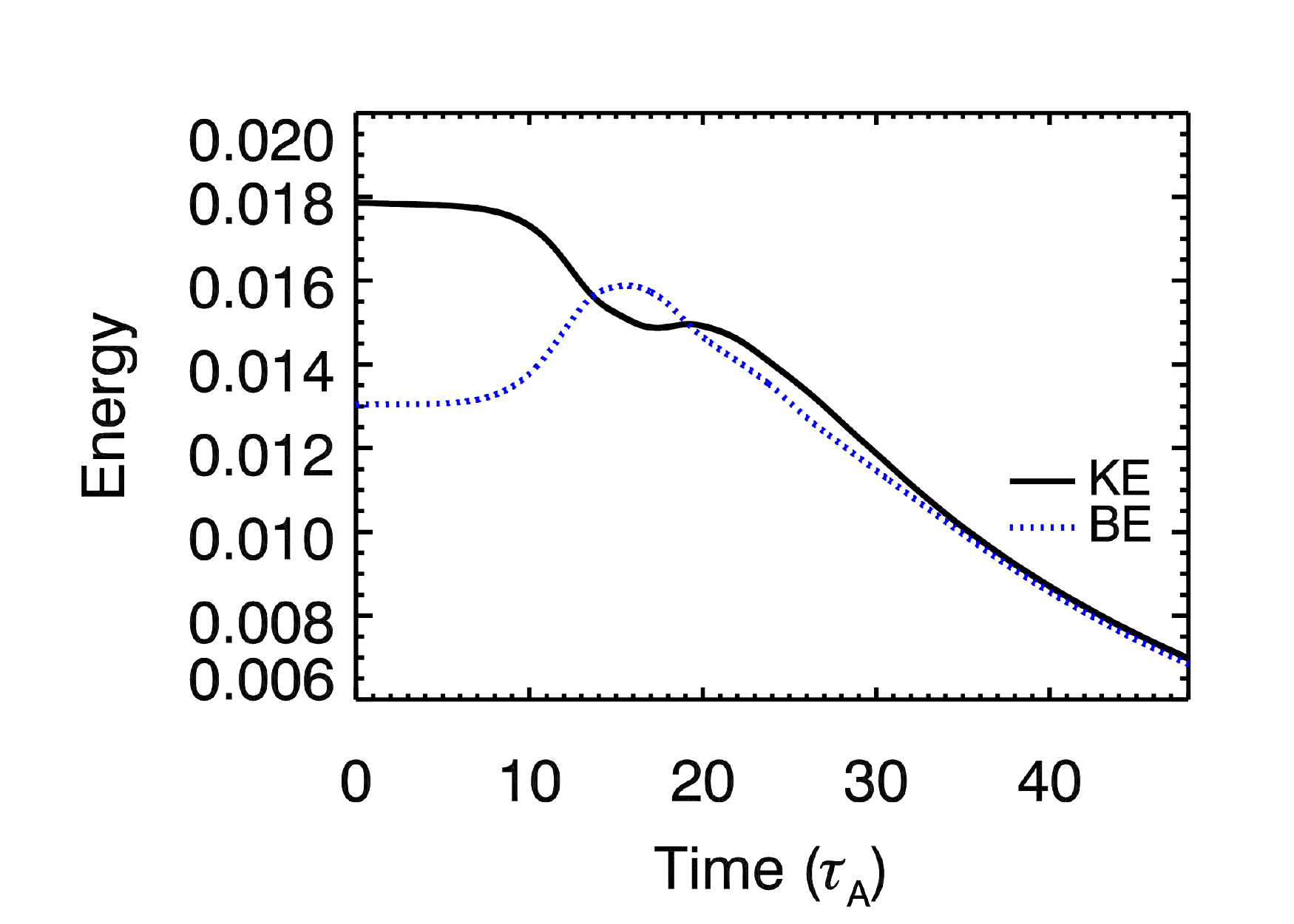}
\includegraphics[type=pdf,ext=.pdf,read=.pdf,width=0.33\textwidth]{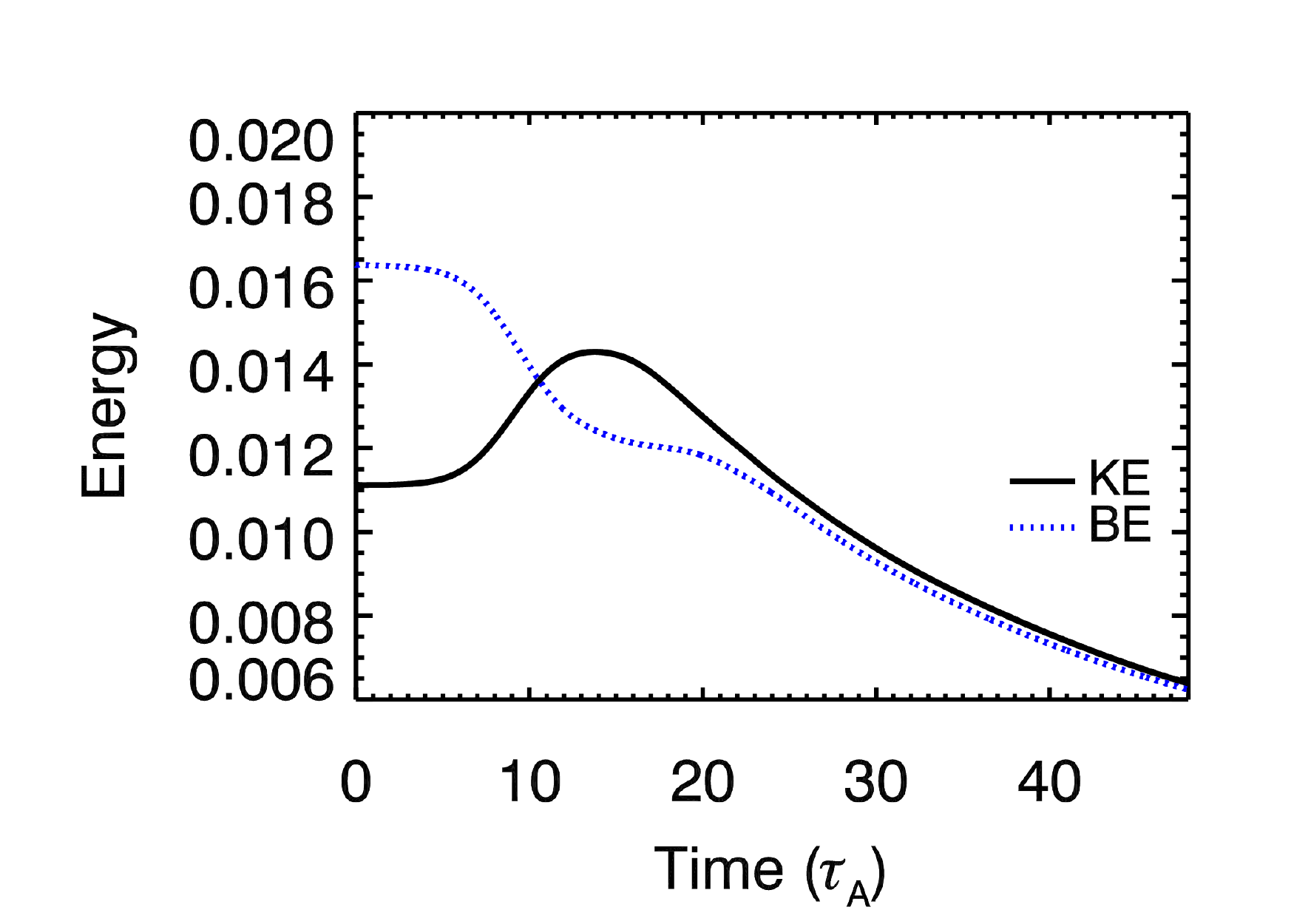}
\includegraphics[type=pdf,ext=.pdf,read=.pdf,width=0.33\textwidth]{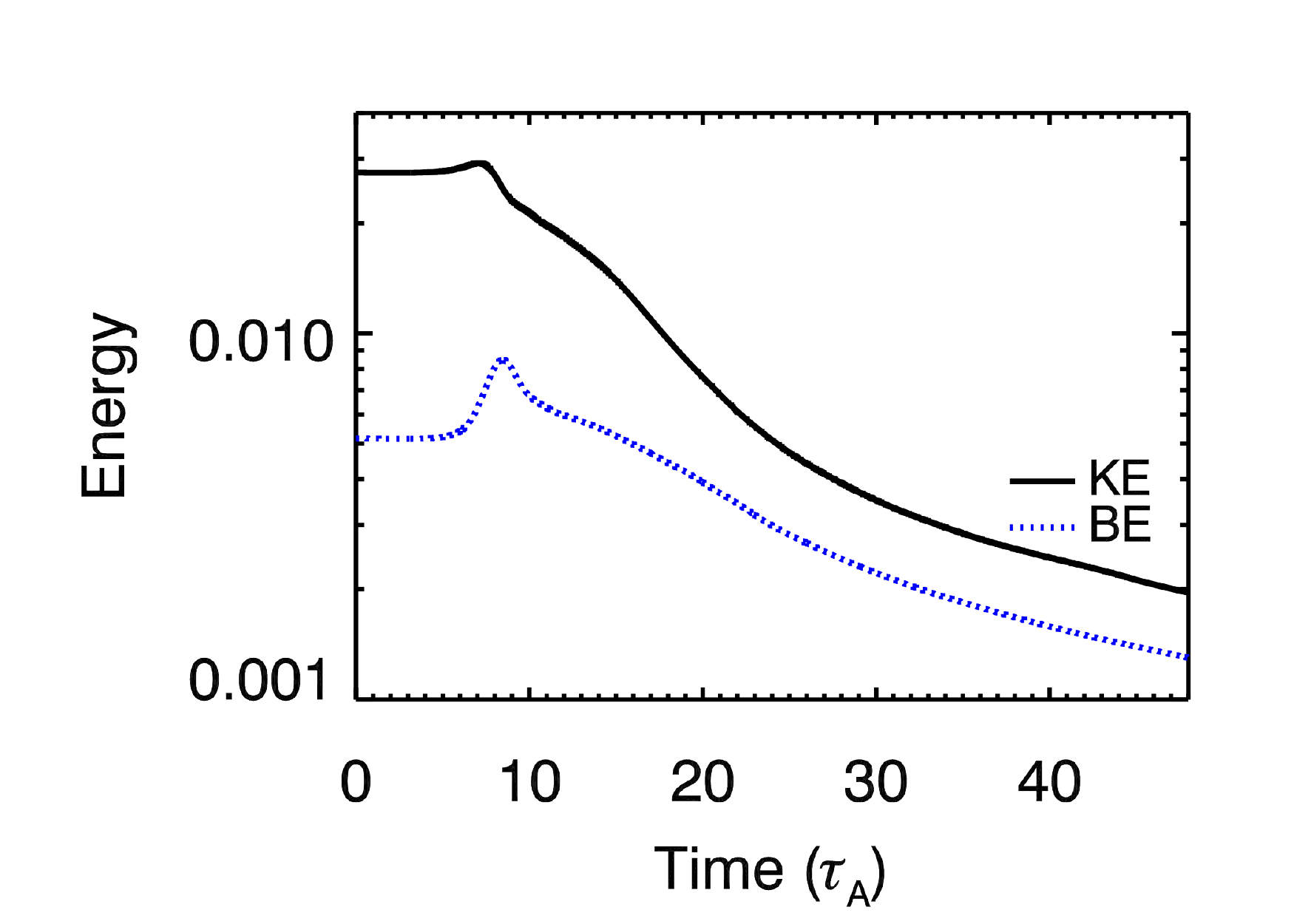}
\end{center}
\caption{Energetics as they depend upon the details of rotation in RPB models, independently normalized so that the total initial energy (TE+BE+KE) is unity.
In model RPB-slow (center), the rotational velocity is halved from the case of the fiducial model RPB (left).
The kinetic and magnetic energies in each run converge to similar values, even though the kinetic/magnetic energies in models RPB-slow and RBP are initially both different and differently ordered.
In model RPB-xrot (right), the rotation profile is extended radially beyond the magnetized region, which lessens this strong connection between the kinetic and magnetic energies.
While all three RPB models succumb to CDI, the global energy profiles depend upon the spatial partitioning of energy in the system.}
\label{fig:RPB_COMPARE_ROT_ENERGY_KB}
\end{figure*}

In Figure \ref{fig:PB_SINGLE_ENERGY_KBT}, we can see the basic pattern of energy exchange that reflects the development of CDI.
Specifically, the total kinetic energy on the grid in model PB increases by approximately four orders of magnitude over several Alfv\'en crossing times.
During that same time period, the magnetic energy begins to decrease.
The kinetic energy increase begins to roll over at $t\sim 6.5~\tau_{\rm A}$, achieving the peak amplitude at $t\sim 7~\tau_{\rm A}$.
After this time, both the kinetic and magnetic energies show a decrease that is attributable primarily to turbulent dissipation into thermal energy and some loss off of the computational boundaries orthogonal to the column axis.
This picture is consistent with the morphological evolution of model PB (Figure \ref{fig:PB_EV}) in which the column bends, increasing the total kinetic energy, and eventually reaches a turbulent state in which the kinetic energy increases are offset by dissipation into thermal energy.
The cyan region of Figure \ref{fig:PB_SINGLE_ENERGY_KBT} depicts the approximate range of time over which the CDI undergo linear growth.
In this region, we measure the maximum growth rate of the instability to be approximately $2.0~\tau_{\rm A}^{-1}$ for the fiducial PB system.

We can compare this growth rate to analytic estimates that have been constructed for physically analogous systems.
\citet{1998ApJ...493..291B}, for example, provides a framework for estimating the growth rate of the toroidal field CDI under very general circumstances.
By inserting the initial conditions of our PB model into this framework, we can compute the expected CDI growth rate on a zone-by-zone basis within our computational grid to find the maximum.
Combining the expressions provided in equations 3.18 and 4.8 of \citet{1998ApJ...493..291B} and inserting our initial values, we find that the maximum expected growth rate in model PB is $\sim 1.9~\tau_{\rm A}^{-1}$.
This is within five percent of the value measured from the energy evolution of our simulation shown in Figure \ref{fig:PB_SINGLE_ENERGY_KBT}.
Moreover, this demonstrates that maximum growth is actually being achieved in a large enough region of the grid that this rate dominates the energetics of the system.

The energy evolution in the fiducial rotating system RPB, on the other hand, is very different than that of model PB even though both columns deform and develop turbulence.
As Figure \ref{fig:RPB_SINGLE_ENERGY_KBT} illustrates, the magnetic energy first experiences a period of growth and is never reduced to the levels seen in model PB.
This is consistent with the visual impression from Figure \ref{fig:RPB_EV} that there are characteristically stronger fields in that model than appear in model PB (Figure \ref{fig:PB_EV}). 
Moreover, after approximately $14~\tau_{\rm A}$, the kinetic and magnetic energies in model RPB follow very similar trends, decreasing to nearly identical values.
As we will soon see, the degree to which these quantities are coupled will depend upon the details of the rotation profile.
In practice, it is challenging to derive an accurate growth rate for CDI in rotating systems from the energy profiles alone since the kinetic energy due to rotation is simultaneously evolving in time.
For that reason, we limit our current discussion of growth rates to the PB and FF classes of simulation.

Finally, the energy evolution of model FF (Figure \ref{fig:FF_SINGLE_ENERGY_KBT}) is grossly similar to that of model PB despite the very different column morphologies that the two models develop.
The primary differences in their energy evolution are that the kinetic energy in model FF experiences a growth of only approximately two orders of magnitude before saturating and that the peak amplitude of the kinetic energy is achieved at a much later time, around $110-130~\tau_{\rm A}$.
Additionally, we note that the growth depicted in Figure \ref{fig:FF_SINGLE_ENERGY_KBT} appears to have two distinct phases, featuring a roughly constant slope from $t=0-35~\tau_{\rm A}$, whereupon it steepens until saturation.
Moreover, the level of kinetic energy always remains a small fraction of the total energy, reaching only a tenth of one percent at maximum.
Thus, we see that CDI do act to increase the kinetic energy in both the FF and PB configurations, but also that the levels to which the kinetic energy is amplified depend strongly upon the model parameters.
The cyan region in Figure \ref{fig:FF_SINGLE_ENERGY_KBT} again marks the period of linear growth.
In the case of the fiducial model FF, the initial growth rate is $\sim 0.04~\tau_{\rm A}^{-1}$, although the maximum growth rate at later times (but prior to saturation) is approximately twice this value.

In the case of an initially force-free field, we can compare this growth rate to that analytically estimated by \citet{2000A&A...355..818A}, for example.
This work provides a very simple closed expression for the growth rate in their Table 1, which is $\omega_{\rm max} \sim 0.13 v_{\rm A}$ for force-free fields of constant pitch angle (identical to those of model FF).
This translates in our case to $\omega_{\rm max} \sim 0.04$, which is identical to the initial growth measured in our simulations.
Again, this is approximately a factor of two less than the maximum growth that is eventually achieved before saturation of the CDI, but it may be that the column evolution changes the Alfv\'en speed sufficiently in portions of the column that higher growth is achievable at later times.
An alternative interpretation is that this second phase of linear growth reflects the development of a different mode number, but the presence of such a mode is not easily inferred from examination of the flow morphology.
It is important to note that both the morphology and linear CDI growth rate of our FF model appear to be very similar to those reported in \citet{2009ApJ...700..684M} for a nearly identical field model, which they refer to as their $\alpha=1$ case.

\subsubsection{The Effects of Initial Conditions}
To explore how the evolution of CDI depends upon the physical parameters in our models, we first examine how the details of rotation contribute to the system energetics in the RPB models.
Specifically, we seek to determine the origin of the increased magnetic energy seen in Figure \ref{fig:RPB_SINGLE_ENERGY_KBT} that seemingly runs contrary to the usual CDI development.
Figure \ref{fig:RPB_COMPARE_ROT_ENERGY_KB} shows a comparison of the fiducial rotating model RPB (left) with models RPB-slow (center), in which $v_{\phi}$ is decreased, and RPB-xrot (right), in which the profile is extended to larger radii.
Model RPB-slow also shows a strong convergence of kinetic and magnetic energies despite the fact that, unlike model RPB, the system was initialized with more magnetic than kinetic energy.
This suggests that the two forms of energy are very likely to co-evolve whenever the CDI become active, somewhat independent of which energy form is dominant.
Although we have not attempted to identify the exact details of this energy transfer, one can imagine a mechanism not unlike the magnetorotational instability \citep{1991ApJ...376..214B} operating to convert rotational energy to magnetic energy once the column has deformed sufficiently that the toroidal field can experience radial shear.
Of course, this presupposes that the two forms of energy are distributed similarly.
The rightmost panel of Figure \ref{fig:RPB_COMPARE_ROT_ENERGY_KB} illustrates that a rotation profile extended beyond the magnetized region leads to parallel, but well-separated energy evolution, as one would expect.

\begin{figure}
\begin{center}
\leavevmode
\includegraphics[type=pdf,ext=.pdf,read=.pdf,width=\columnwidth]{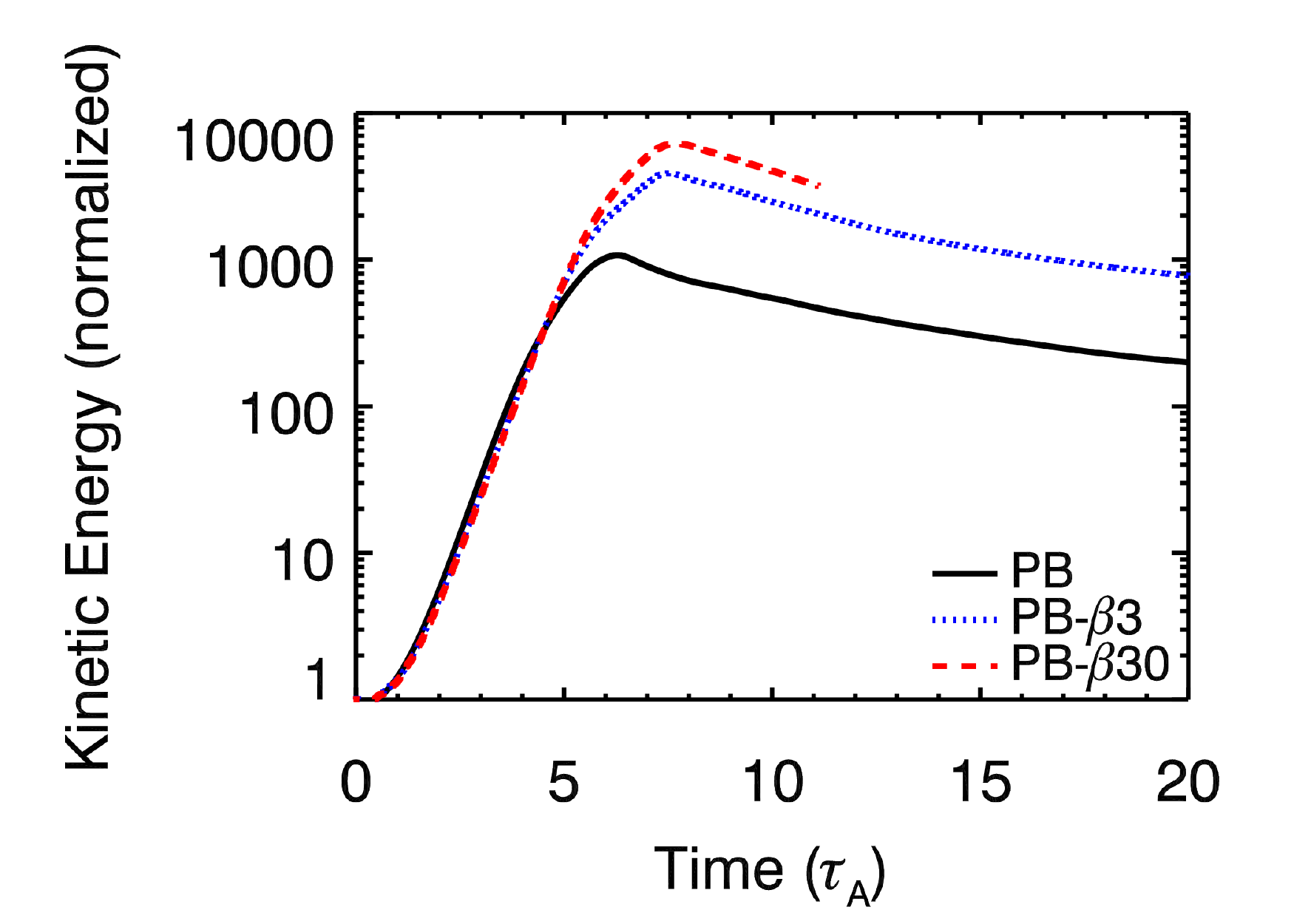}
\end{center}
\caption{The evolution of kinetic energy over time as a function of $\beta$ for models PB (solid line), PB-$\beta$3 (dotted line), and PB-$\beta$30 (dashed line).  For each model, the initial kinetic energy is independently normalized to unity.  Scaling the evolution by the model-dependent Alfv\'en time ($\tau_{\rm A}$) illustrates that the linear growth rate of CDI is relatively model independent, although the magnitude of the total amplification and turnover time both increase with increasing $\beta$.}
\label{fig:PB_COMPARE_BETA_ENERGY_K}
\end{figure}

\begin{figure}
\begin{center}
\leavevmode
\includegraphics[type=pdf,ext=.pdf,read=.pdf,width=\columnwidth]{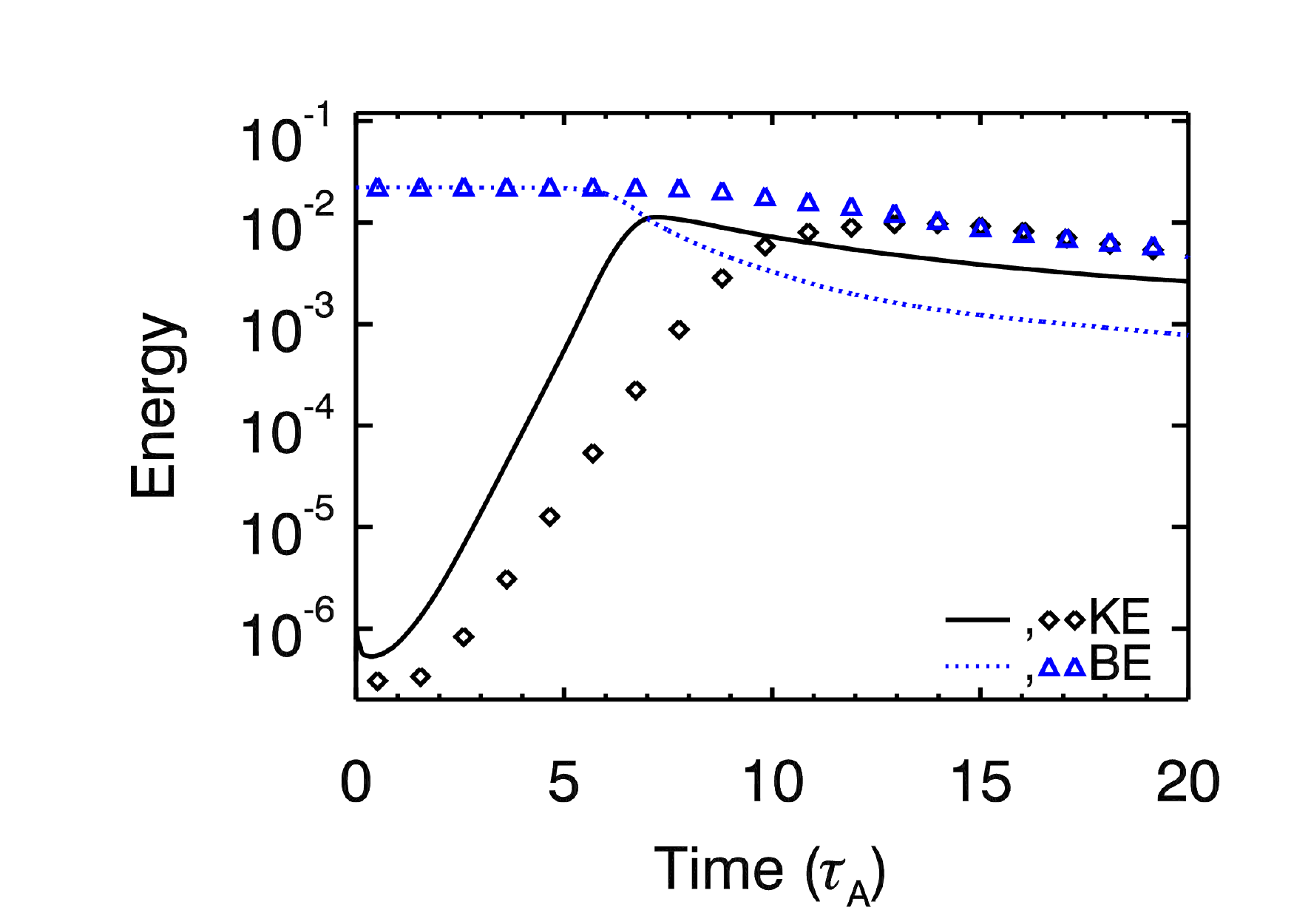}
\end{center}
\caption{Comparing the kinetic and magnetic energy evolution between the fiducial PB model (lines) with model PB-kom (glyphs), which features a Komissarov-type field profile.  Each model is independently normalized so that the total initial energy (TE+BE+KE) is unity.  While both models are disrupted by the CDI, the details of the energetics, including the CDI growth rate, differ according to the initial field geometry and energy distributions.  The linear growth rates are comparable to those predicted analytically for the two configurations.}
\label{fig:PB_COMPARE_KOMISS_ENERGY_KB}
\end{figure}

We are also interested in determining how CDI development is affected by magnetic field strength as parameterized by the plasma $\beta$.
Figure \ref{fig:PB_COMPARE_BETA_ENERGY_K} shows how the kinetic energy evolution in the PB series of models varies with three different field strengths, reflecting both subthermal and superthermal configurations.
In all cases, the initial perturbation was set to the same (small) fraction of the initial maximum Alfv\'en speed, meaning that the initial kinetic energies differed between models.
We have therefore independently normalized the initial kinetic energies to unity so that we may more easily cross-compare the models.
Clearly, Figure \ref{fig:PB_COMPARE_BETA_ENERGY_K} illustrates that the linear growth stages of the CDI are not dramatically affected by the initial field strength.
The growth is not only parallel but nearly identical for each model until approximately $5~\tau_{\rm A}$, where the models begin to differentiate from one another.
Specifically, the maximum saturation level and saturation time both appear to increase with increasing $\beta$ (\ie reduced field strengths).

Additionally, we would like to know the importance of the magnetic field profile in determining the evolution of the CDI.
Figure \ref{fig:PB_COMPARE_KOMISS_ENERGY_KB} shows the evolution of the alternate Komissarov-type magnetic field that is counterbalanced by a gas pressure gradient.
Both runs clearly show evidence of kinetic energy enhancement as a result of CDI. 
As measured from the kinetic energy evolution, the CDI growth rates in the two models are comparable but not identical, suggesting that the exact distributions of field and gas pressure do play a role in determining the system evolution, as the analytic theory presented in \citet{1998ApJ...493..291B} suggests.
The growth rate in model PB-kom is measured to be $\sim 1.39~\tau_{\rm A}^{-1}$, which is about $70\%$ that of model PB.
If we again model the expected growth rate based on the expressions provided in \citet{1998ApJ...493..291B}, we anticipate a growth of $\omega \sim 1.27~\tau_{\rm A}^{-1}$, which is within $10\%$ of what is measured.
The other significant difference introduced by our choice of field structure is that PB-kom achieves equipartition between magnetic and kinetic energies while the kinetic energy in model PB evolves to exceed equipartition.
We should avoid over-interpreting this feature, however, since the global equipartition shown in the figure will not necessarily be indicative of local energy distributions everywhere on the grid.
Specifically, one can easily imagine that the different field distributions in the two models lead to different total magnetic energies even if both systems experience rather similar development of CDI.
In fact, direct inspection of the data confirms that there are a few local regions in model PB-kom where the kinetic and magnetic energies differ by a few orders of magnitude even though this is not reflected in the global average.

\begin{figure}
\begin{center}
\leavevmode
\includegraphics[type=pdf,ext=.pdf,read=.pdf,width=\columnwidth]{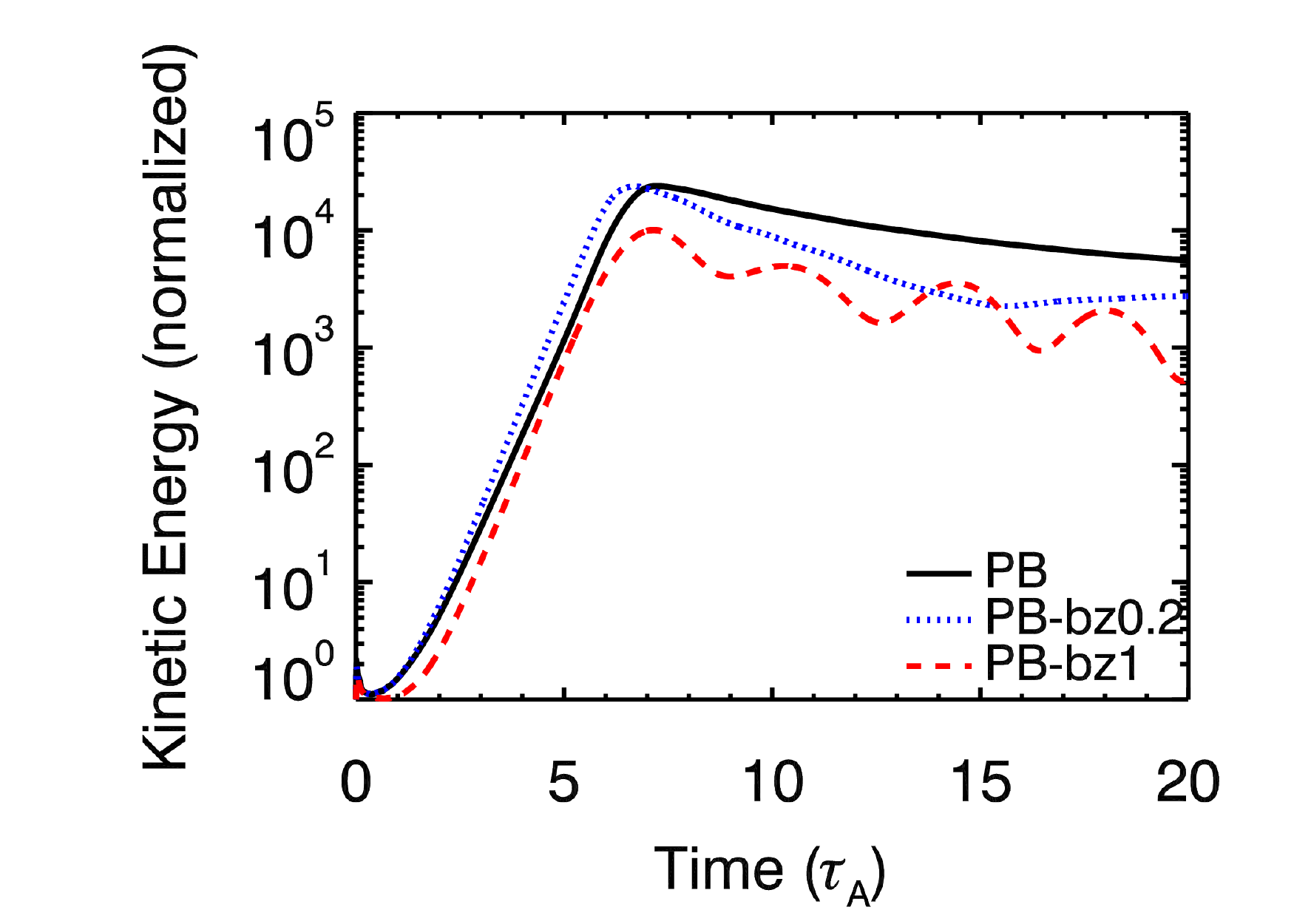}
\end{center}
\caption{Comparing the kinetic energy evolution between the fiducial PB model (line) with models featuring poloidal fields that are comparable to $100 \%$ (PB-bz1) and $20\%$ (PB-bz0.2) of the toroidal field strength.  Each model is independently normalized so that the initial kinetic energy is unity.  The linear growth of the CDI is not dramatically affected, although the maximum kinetic energy achieved and saturation behaviors are dramatically different.}
\label{fig:PB_COMPARE_BZ_ENERGY_K}
\end{figure}

Considering that there are significant morphological and energetic differences between the FF and PB models, it is important to examine the role of poloidal field in the evolution of these systems.
Specifically, we seek to determine whether the presence of poloidal field is sufficient to retard the development of CDI when introduced into the PB class of models.
Figure \ref{fig:PB_COMPARE_BZ_ENERGY_K} shows the kinetic energy evolution for models PB-bz1 and PB-bz0.2, in which $|B_z| = |B_{\phi}|$ and $|B_z| = 0.2|B_{\phi}|$, respectively.
These models differ from all of the others in that the magnetic pitch of these field configurations is proportional to the radius, rather than being constant.
First, we note that the linear growth of CDI is quite similar in all cases, confirming that the different growth rates seen between the FF and PB models cannot be attributed to the presence of a strong poloidal field in model FF.
That said, it is clear that the linear growth terminates early for PB-bz1, suggesting that, as expected, poloidal fields do play a major role in the saturation of the CDI.
Furthermore, the saturation levels of the instability decrease on average with increasing poloidal field strength, although with substantial fluctuations that are of order a few times the characteristic Alfv\'en crossing time.
Incidentally, the column morphologies that develop for the poloidal models are also intermediate between those of the strongly turbulent PB model and the less disordered FF class of simulation.

\begin{figure}
\begin{center}
\leavevmode
\includegraphics[type=pdf,ext=.pdf,read=.pdf,width=\columnwidth]{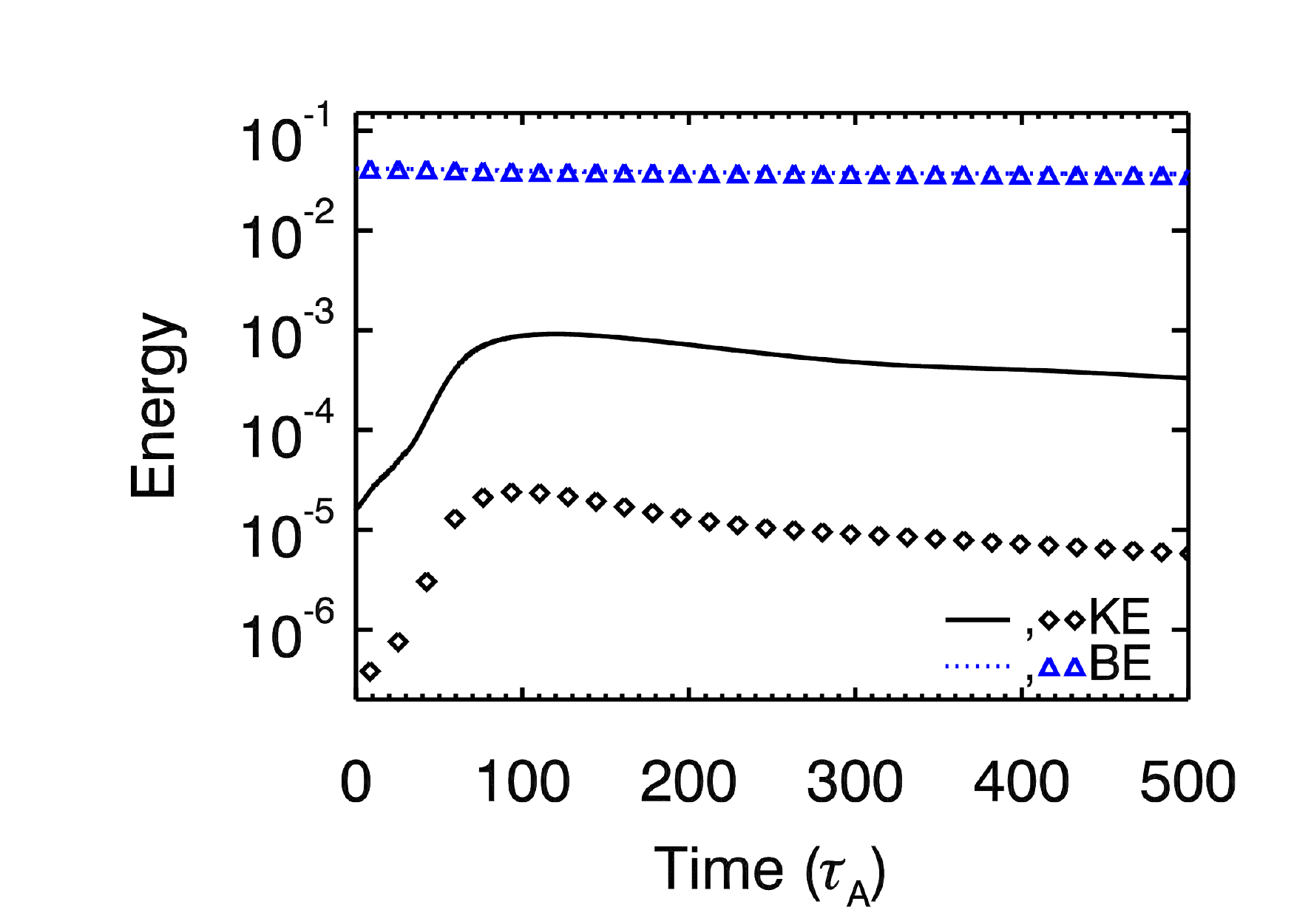}
\end{center}
\caption{Comparing the kinetic and magnetic energy evolution between the fiducial FF model (lines) and model FF-hot (glyphs), in which the gas is relativistically hot (\ie $p \sim 10 \rho$).  Each model is independently normalized so that the total initial energy (TE+BE+KE) is unity.  The two models feature similar CDI growth rates and parallel long-term evolution.}
\label{fig:FF_COMPARE_HOT_ENERGY_KB}
\end{figure}

Next, we explore the degree to which the temperature of the column affects its evolution.
Model FF-hot, in which the gas pressure is a factor of ten greater than the rest mass energy, is unique among our models in that special relativity is essential for achieving a subluminal sound speed, even in the initial conditions.
Figure \ref{fig:FF_COMPARE_HOT_ENERGY_KB} compares the evolution of kinetic and magnetic energy in FF-hot with the fiducial FF model.
Again, both models feature clear indications of the development and saturation of the CDI.
In fact, the energy evolution in model FF-hot runs roughly parallel to that of the fiducial model, with a displacement that reflects the different initial fractional distributions of energy.
This suggests that the temperature of the gas, even as it enters into the properly relativistic regime, does not strongly affect the development of the CDI.

\begin{figure*}[b]
\begin{center}
\leavevmode
\includegraphics[type=pdf,ext=.pdf,read=.pdf,width=0.49\textwidth]{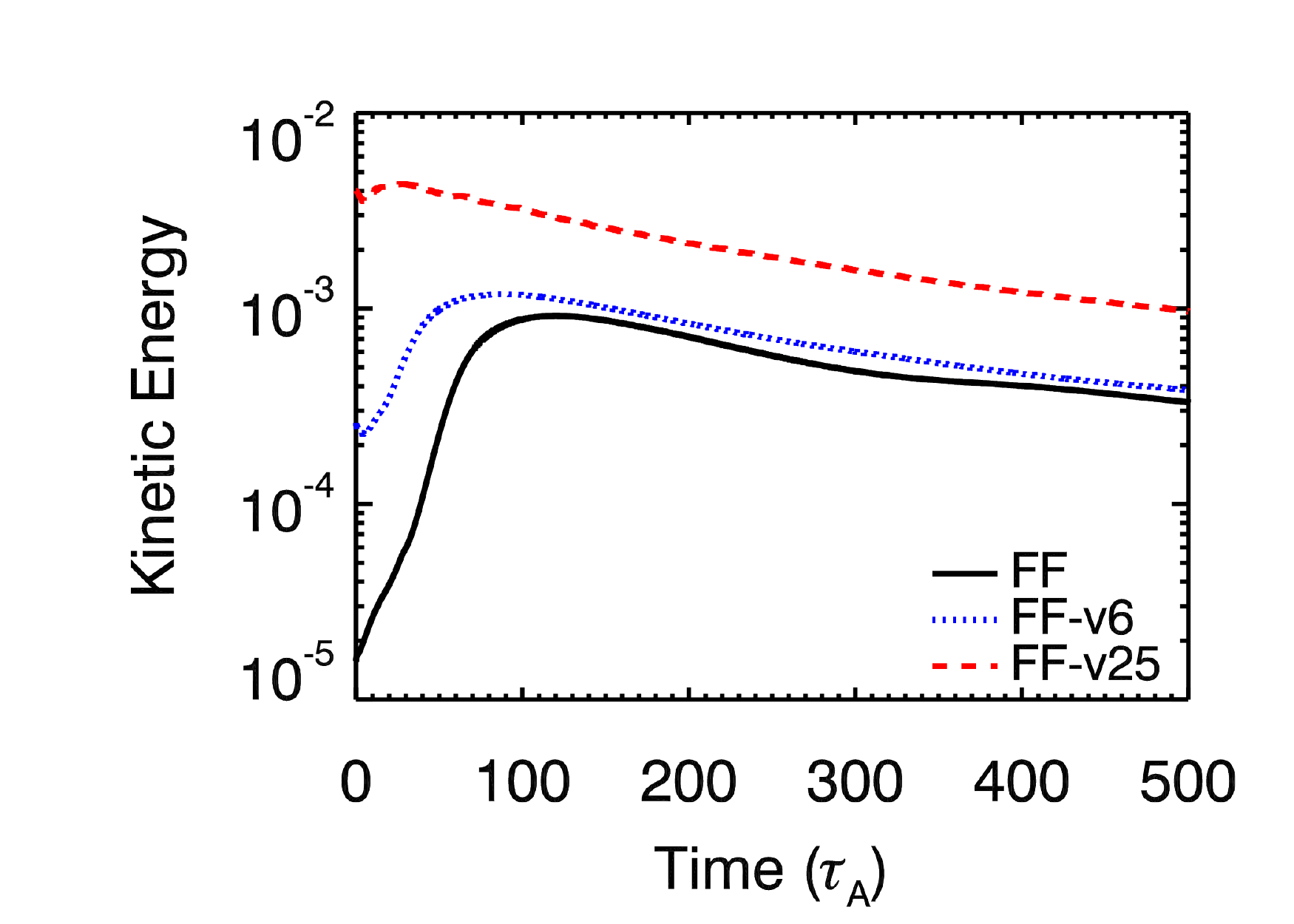}
\includegraphics[type=pdf,ext=.pdf,read=.pdf,width=0.49\textwidth]{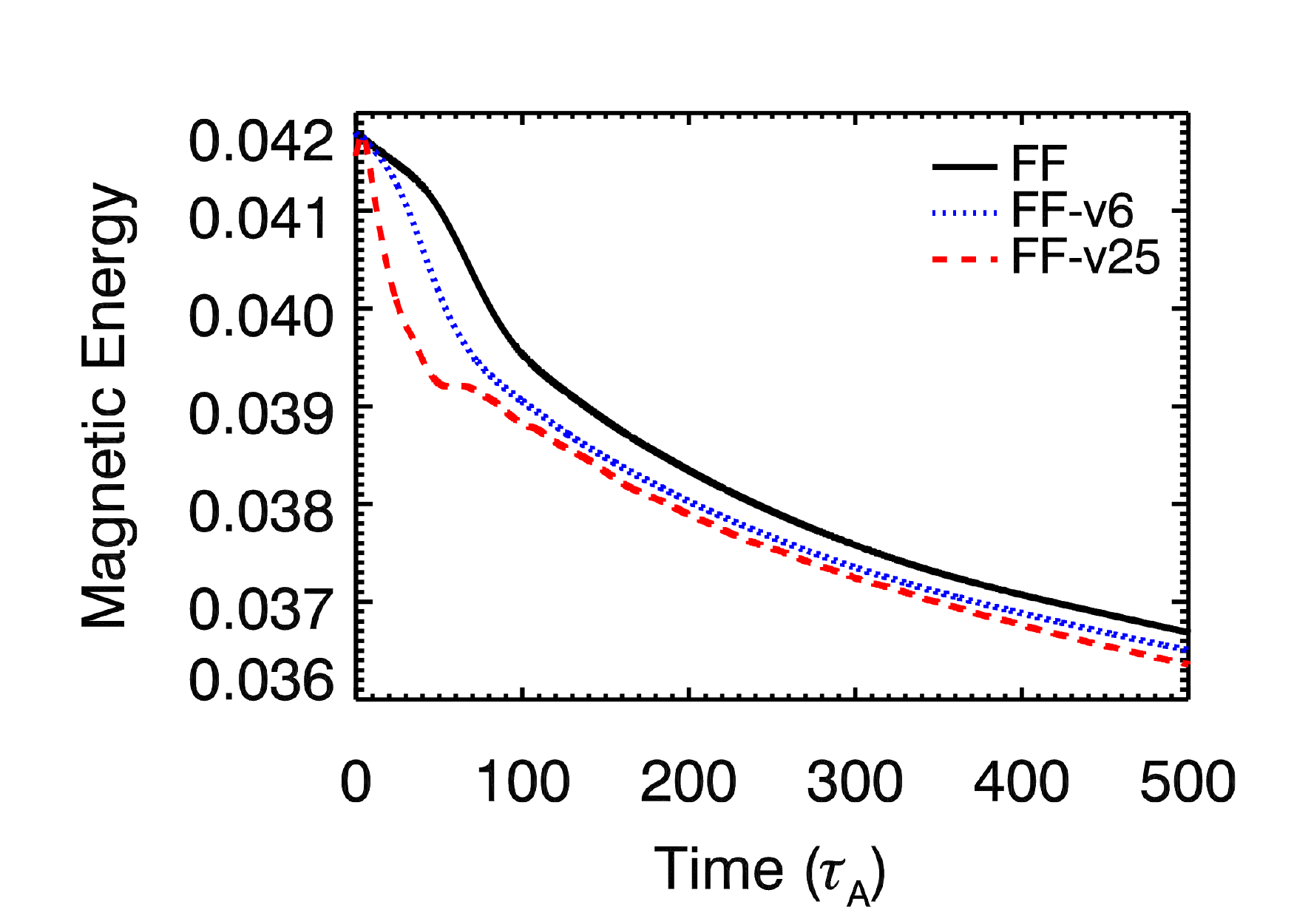}
\end{center}
\caption{Energy evolution in the FF models as it depends upon perturbation magnitude for kinetic energy (left) and magnetic energy (right).  Each model is independently normalized so that the total initial energy (TE+BE+KE) is unity.  Larger perturbations (\eg FF-v25) obviously lead to larger initial/final kinetic energies, but all perturbations produce roughly parallel evolutions at late times.}
\label{fig:FF_COMPARE_KICK_ENERGY_KB}
\end{figure*}

Finally, we examine the influence of the applied perturbation.
The perturbation used in the vast majority of our models is a large-scale non-axisymmetric kick designed to excite the $|m|=1$ mode of the CDI.
Generally, this kick has a strength of approximately one percent of the initial maximum Alfv\'en speed so that the system is excited, but not initially pushed too far away from the equilibrium state.
Still, it is worth determining whether or not the exact magnitude of this kick affects the evolution of CDI in a significant manner.
Figure \ref{fig:FF_COMPARE_KICK_ENERGY_KB} depicts the energy evolution of initially force-free models in which the kicks were $6\%$ (FF-v6) and $25 \%$ (FF-v25) of the maximum Alfv\'en speed in otherwise identical simulations.
The most obvious difference between these models naturally appears in the overall kinetic energy levels, which reflect the higher initial kinetic energies associated with larger kicks.
Otherwise, the evolution of CDI, particularly in the non-linear phases, runs roughly parallel for each different model, suggesting that the precise value of the kick is not particularly important for the evolution of the system.

\begin{figure*}
\begin{center}
\leavevmode
\includegraphics[type=pdf,ext=.pdf,read=.pdf,width=0.49\textwidth]{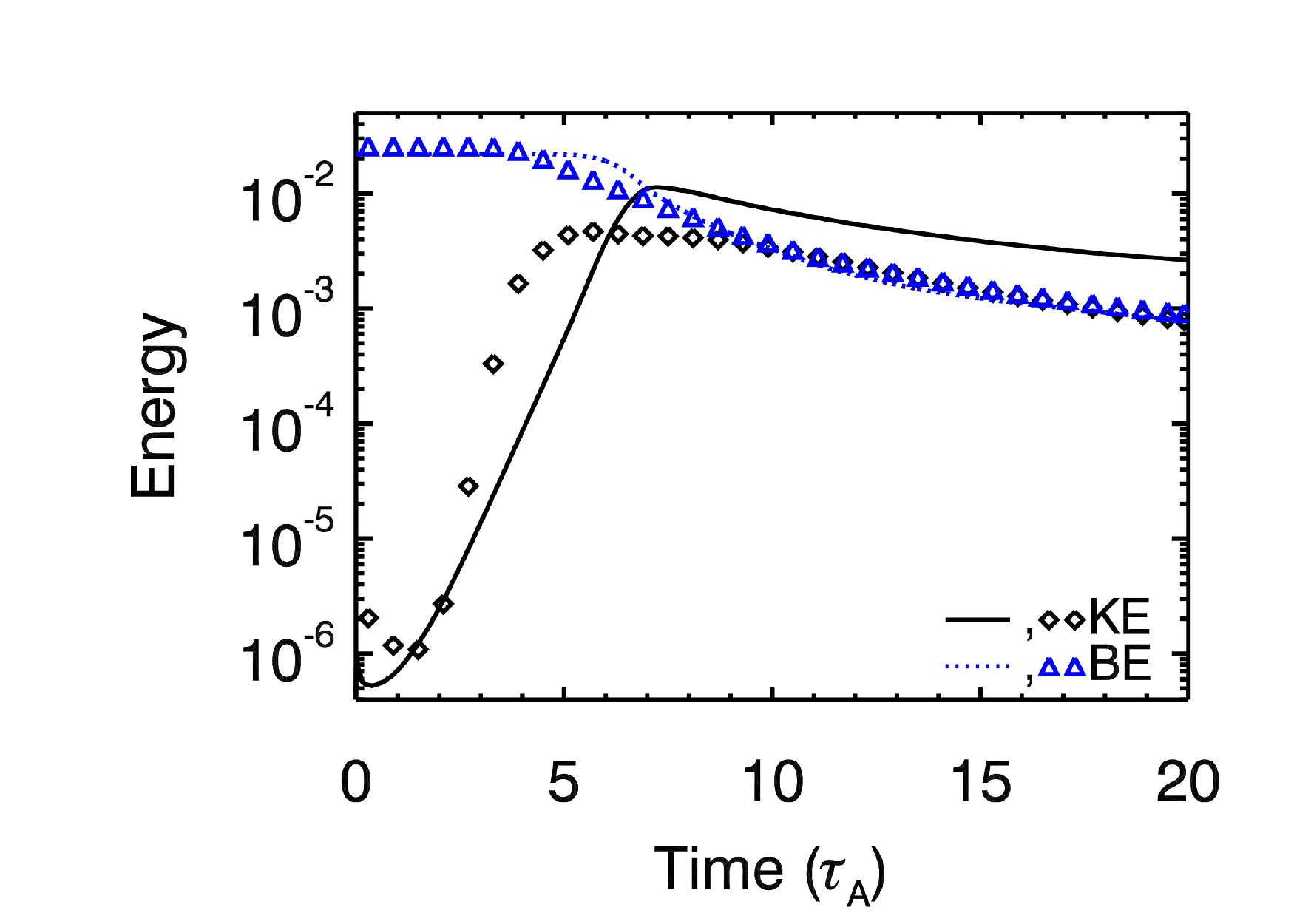}
\includegraphics[type=pdf,ext=.pdf,read=.pdf,width=0.49\textwidth]{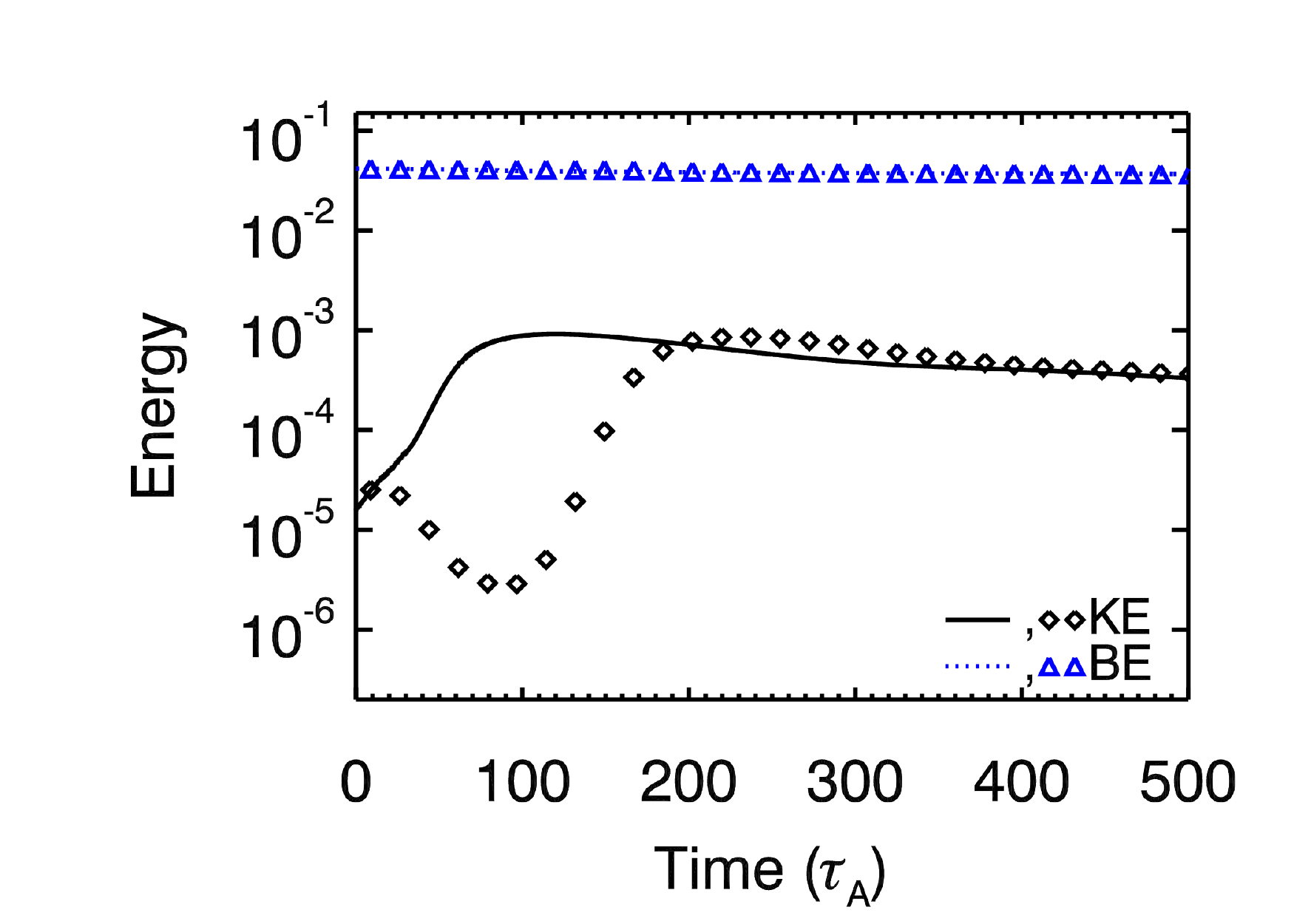}
\end{center}
\caption{Energy evolution in the PB (left) and FF (right) models, comparing whether or not the system is subjected to an organized $|m|=1$ perturbation (lines) or a random perturbation (glyphs).  Each model is independently normalized so that the total initial energy (TE+BE+KE) is unity.  Both models are obviously subject to the growth of CDI from random perturbations, but the PB model is more affected by the details of the perturbation than is the FF model, which is simply displaced in time.}
\label{fig:FF_COMPARE_RANDPERT_ENERGY_KB}
\end{figure*}

One may also explicitly test to determine whether the $|m|=1$ mode is really the CDI mode that we should be exploring.
Linear theory tells us that this is under many conditions the fastest growing mode (\eg \citealt{1998ApJ...493..291B}), but it is important to check that this mode is naturally excited by a more generic perturbation than is employed in most of our models.
Figure \ref{fig:FF_COMPARE_RANDPERT_ENERGY_KB} compares the fiducial PB and FF models with their randomly perturbed analogues, PB-vrand (left) and FF-vrand (right).
Here, the random perturbations are generated by using a random number generator to produce a random (in strength and orientation) velocity in each computational zone such that the RMS velocity magnitude as measured over the entire grid is comparable to the maximum kick applied in the fiducial models.
This is an admittedly simple test that would not, for example, be expected to converge numerically for increasing resolution, given that the perturbations are applied at the grid scale.
Still, if CDI develop on large scales naturally from this type of perturbation, we can rest assured that our simulations reflect a fairly robust set of generic outcomes.

In Figure \ref{fig:FF_COMPARE_RANDPERT_ENERGY_KB}, we see the telltale sign of CDI development, namely a clear increase in the kinetic energy on the grid.
There are some interesting points to note, however.
First, the PB and PB-vrand models (right) achieve rather similar magnetic energy evolution, but different kinetic energy progressions.
Specifically, the kinetic energy in model PB-vrand demonstrates a slightly different growth rate and saturation level from that of the fiducial model PB.
This is not entirely unexpected since the velocity field has time to modify the pressure (and field) profiles, both of which can affect the CDI growth rate in pressure-supported columns, before the CDI have finished growing.
Additionally, model PB-vrand achieves approximate equipartition between magnetic and kinetic energies at saturation, which is different from the PB model for which the kinetic energy eventually exceeds the magnetic energy globally.
In the FF and FF-vrand models, on the other hand, we see that the kinetic energies evolve quite similarly, which is consistent with the fact that the analytic growth rate estimate for force-free columns includes only the characteristic Alfv\'en speed and radius, neither of which should change dramatically as a result of a random perturbation.
The only significant difference is an offset in time, but this is only natural given that the $|m|=1$ mode was not specifically excited.

\subsubsection{Including Alternative Physics}

\begin{figure}
\begin{center}
\leavevmode
\includegraphics[type=pdf,ext=.pdf,read=.pdf,width=\columnwidth]{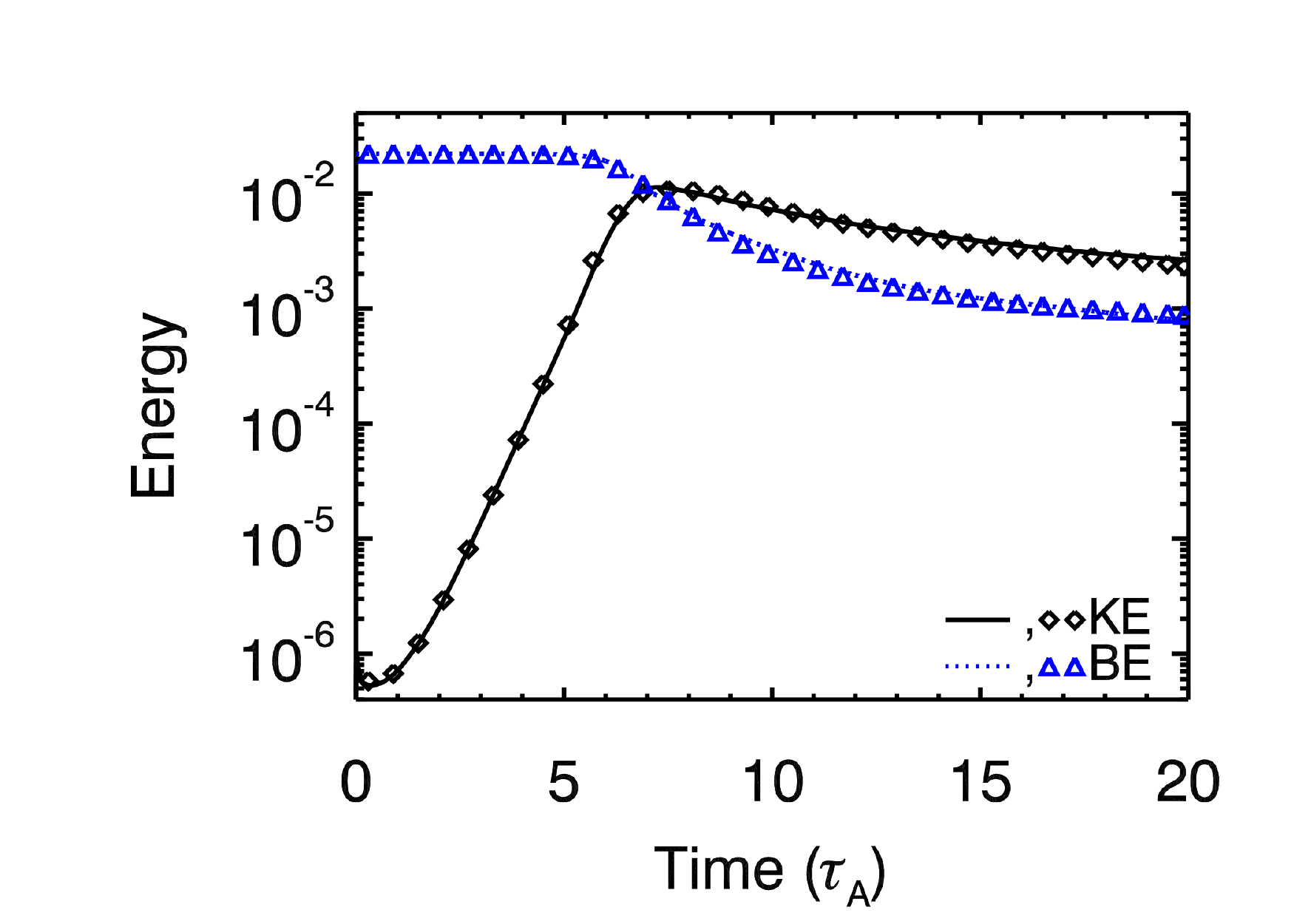}
\end{center}
\caption{Comparing the effects of the value of the adiabatic index between the fiducial model PB (lines) with $\gamma=5/3$ and model PB-g1.33 (glyphs), in which $\gamma=4/3$.  Each model is independently normalized so that the total initial energy (TE+BE+KE) is unity.  The model evolution is fairly insensitive to this parameter, with maximum deviations of approximately $10\%$ in the kinetic and magnetic energies by the end of the evolution.}
\label{fig:PB_COMPARE_ADIABATIC_ENERGY_KB}
\end{figure}

In addition to exploring the influence of our choice of model parameters, we also seek to determine the significance of the included physics.
For example, although it is not obvious that these systems experience a significant amount of compression as they evolve, it is worth determining whether a change of the adiabatic index has any noticeable effects.
Figure \ref{fig:PB_COMPARE_ADIABATIC_ENERGY_KB} compares the evolution of fiducial model PB with that of PB-g1.33 in which the adiabatic index of 4/3 corresponding to a relativistic fluid is employed.
There are essentially no important differences between these two models beyond an approximately $10\%$ deviation in the amount by which the kinetic energy is amplified after 20 Alfv\'en crossing times.
This is a minor enough discrepancy that we can safely conclude that the value of the adiabatic index does not play a major role in the evolution of CDI.
One could imagine also investigating the effects of a variable Synge-type \citep{synge1957} equation of state, but the relatively small ranges of temperature (a factor of 10, typically) achieved in a given CDI simulation suggest that this would not have a major effect on these models.

A more fundamental physical difference may be expected to result from the incorporation of special relativistic physics.
While our models are always initialized with low fluid velocities, the Alfv\'en speeds in these magnetized columns can be quite large.
Specifically, the fiducial models all feature maximum $v_{\rm A} > 0.1$, with the FF model having initial values as high as $v_{\rm A} = 0.32$.
This is sufficiently large that it is not obvious how significant a role special relativity will play in regulating this speed as the flow evolves.
Furthermore, it is unclear whether the fluid velocities that develop as a result of CDI will be high enough to be strongly affected by the natural speed limit imposed by special relativity.

\begin{figure}
\begin{center}
\leavevmode
\includegraphics[type=pdf,ext=.pdf,read=.pdf,width=\columnwidth]{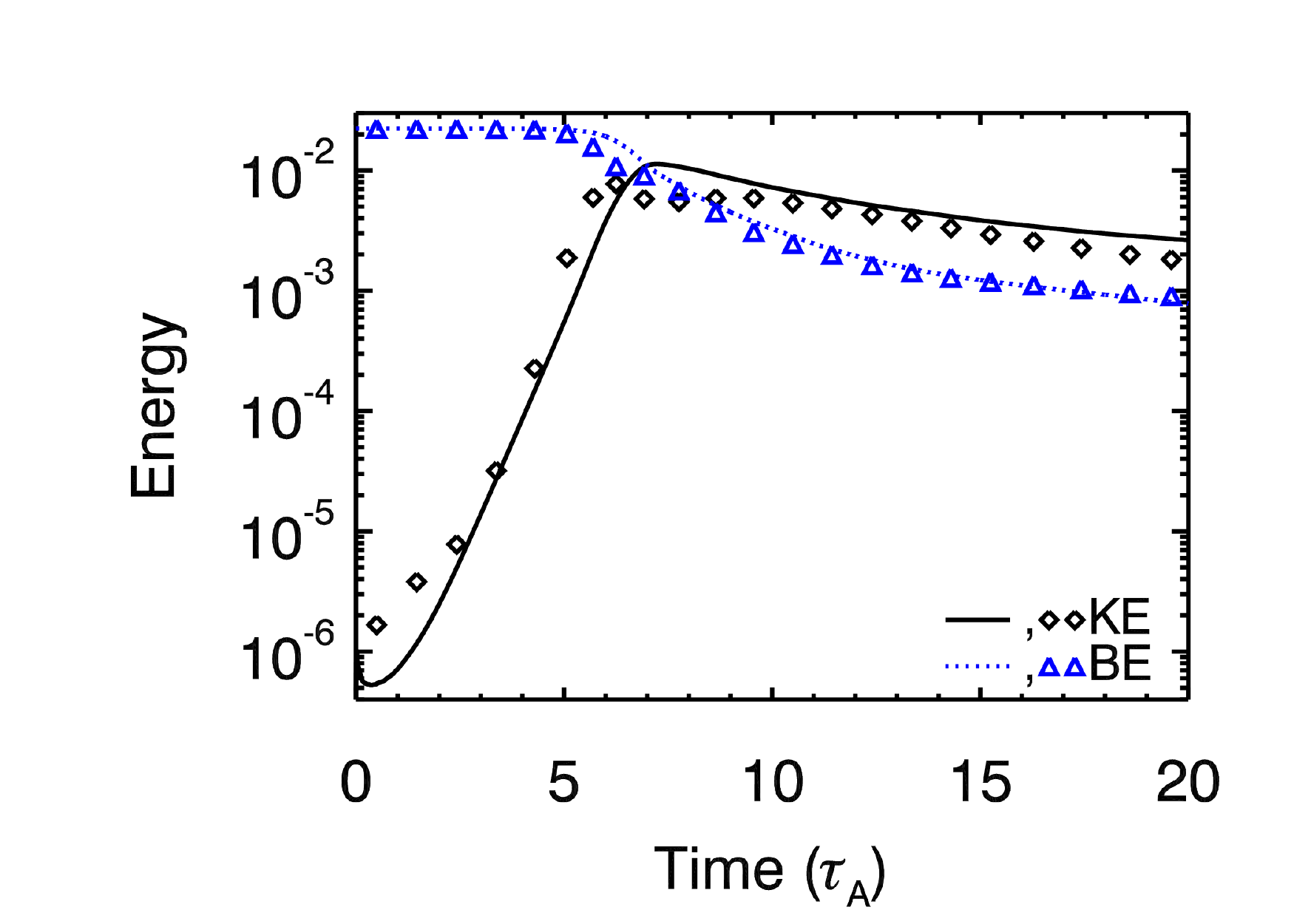}
\end{center}
\caption{A comparison between the fiducial model PB (lines) and its Newtonian analog, model PB-newt (glyphs).  Each model is independently normalized so that the total initial energy (TE+BE+KE) is unity.  The two models evolve in a similar fashion, although deviations can be seen, particularly in the kinetic energy where the relativistic value is ultimately $50\%$ greater than that of the Newtonian case.}
\label{fig:PB_COMPARE_NEWT_ENERGY_KB}
\end{figure}

\begin{figure}
\begin{center}
\leavevmode
\includegraphics[type=pdf,ext=.pdf,read=.pdf,width=\columnwidth]{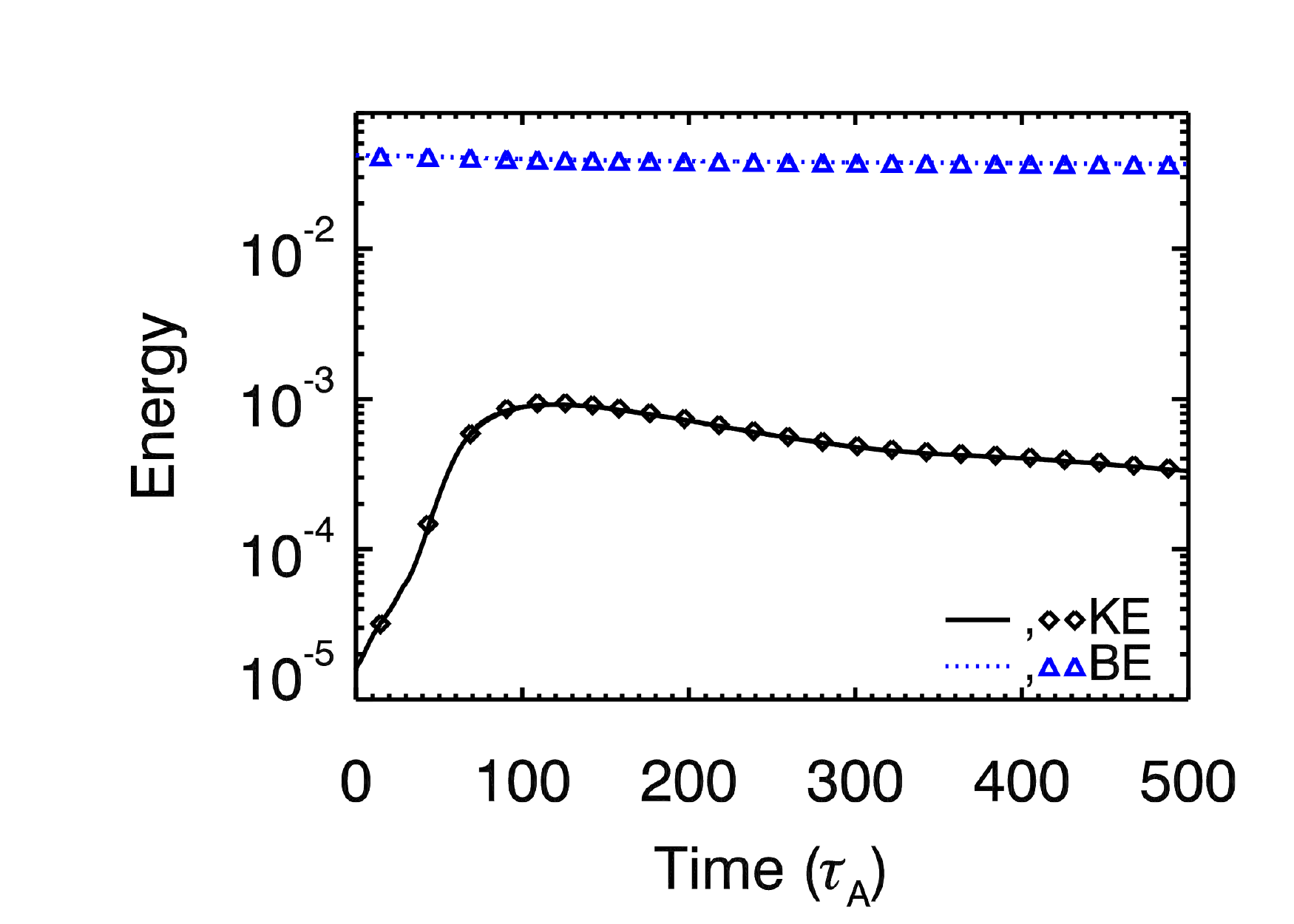}
\end{center}
\caption{A comparison between the fiducial model FF (lines) and its Newtonian analog, model FF-newt (glyphs).  Each model is independently normalized so that the total initial energy (TE+BE+KE) is unity.  The FF models evolve more similarly than the PB models, with only a few percent differences between relativistic and Newtonian physics.}
\label{fig:FF_COMPARE_NEWT_ENERGY_KB}
\end{figure}

Figures \ref{fig:PB_COMPARE_NEWT_ENERGY_KB} and \ref{fig:FF_COMPARE_NEWT_ENERGY_KB} show the evolution of the PB-newt and FF-newt models, respectively.
It is initially surprising that the FF models are actually less sensitive to the inclusion of special relativity than are the PB models, despite their nominally larger initial maximum Alfv\'en velocities.
Direct examination of the data reveals that this is likely attributable to the larger flow velocities achieved in the PB models.
When the PB columns are disrupted, small regions can attain velocities of up to nearly 0.14c.
This is three times the maximum velocity reached in the FF models and sufficient to be affected by the presence/absence of relativity.
As a result, the fiducial PB model features a final kinetic energy that is nearly 1.5 times that of model PB-newt, while the FF models differ by only a few percent.
This demonstrates that relativity does play a role in the evolution of these models, but also that special relativistic effects do not dominate the energetics, as might be expected in simulations featuring relativistic shear layers or rotation profiles.

\section{Discussion and Conclusions}
\label{sec:conclusions}
We have reported on an ensemble of local co-moving jet simulations designed to explore the role of current-driven instabilities in magnetized plasma columns.
Our most significant discovery is that the details of initial force balance have a dramatic impact on the resulting column morphology.
Despite the fact that the force-free, pressure-supported, and rotation-supported magnetized columns under consideration are all nominally unstable to CDI, they clearly develop structures that are very dissimilar from one another.
Specifically, our force-free columns undergo a relatively modest degree of deformation that slows dramatically when the CDI achieve non-linear saturation.
The pressure-balanced columns, on the other hand, shred and develop turbulent structures that facilitate continued mixing throughout the evolution of the system.
While the inclusion of rotation in pressure-supported systems adjusts some of the details, these models too are characteristically unstable and undergo significant mixing resulting from the combination of CDI and radial shear.

Despite these grossly different morphologies, the general pattern of CDI energy evolution is similar for all models.
All systems experience a linear growth phase of the CDI, which we have found matches analytic predictions with a high degree of accuracy.
The force-free fields generally experience much slower rates of CDI development and less kinetic energy amplification than their pressure-supported counterparts.
Perhaps the most interesting energy exchange occurs in the rotating systems, where mixing ensures that spatially co-located kinetic and magnetic energies will evolve to approximate equipartition, regardless of their initial ordering.
Considering that most astrophysical winds and jets should possess some amount of angular momentum, with AGN jets being particularly likely to feature very rapid rotation near their source, this largely unrecognized channel of energy exchange could potentially be very important.

Moreover, we have shown that these behaviors are fairly generic outcomes for a wide range of physical and numerical parameters.
The strength and/or structure of the magnetic field, for example, makes much less difference than whether the initial force balance was achieved through force-free fields, pressure gradients, or rotation.
Introducing a poloidal field component into a pressure-supported configuration does not affect the linear growth rate of CDI, although it does eventually adjust the non-linear saturation properties.
The adiabatic index or neglect of special relativity makes only a minor difference, although we have yet to merge special relativity and rotation in a single model.
One expects that special relativity will play a far greater role in affecting flows bounded by a strong shear layer and/or rotating at relativistic speeds.

Numerical considerations prove to be some of the most significant factors in determining the non-linear states of these systems.
Although the choice of Riemann solver and variable reconstruction order do not strongly impact the linear growth phase of CDI, the peak kinetic energy attained and saturation levels are affected by algorithmic choices to a degree beyond most of the physical parameters that we varied.
This serves as a useful reminder that not all numerical approaches are equally applicable to all problems and, specifically, that the HLLE solver should be avoided for problems in which it is important to resolve contact and rotational discontinuities.
Given that the effective resolution of HLLD is greater than that of HLLE (\eg \citealt{2009MNRAS.393.1141M}), this is likely to be even more of an issue for global jet simulations that typically employ lower physical resolutions over larger domains than their local counterparts.

The different methods of force balance correspond to distinct regions of astrophysical parameter space.
Force-free field configurations in pulsar winds or astrophysical jets, for example, are most likely to occur nearest to points of origin.
Simple arguments about flux conservation suggest that toroidal field will eventually dominate over poloidal field in any geometrically expanding jet or asymmetric wind \citep{1984RvMP...56..255B}.
Furthermore, particles are more likely to be heated by dissipation downstream of their origin, with baryons in particular retaining this heat more easily than electrons or positrons.
This provides a convergence of evidence that pressure-supported models are more relevant at greater distances from the source.

That AGN jets are observed to be collimated over many orders of magnitude in physical scale, however, suggests that the disruptive behaviors seen in our pressure-supported and rotating columns do not completely dissociate real, astrophysical systems.
One obvious difference between our current models and astrophysical jets is the presence of a confining medium that forms a shear layer with the jet.
If nothing else, such a layer would certainly interact with and modify any CDI modes that intersected it, as has been seen in simulations by \citet{2002ApJ...580..800B} and \citet{2011ApJ...734...19M}, for example.
Under some circumstances, the shear layer would also be expected to develop KHI modes that would have the potential both to disrupt the shear layer and to interact with the existing CDI modes.
It is also possible, as has been suggested in the weak-field limit by \citet{2002ApJ...580..800B}, that the KHI and CDI modes conspire to stabilize the jet against disruption.
It may even be that the details of jet rotation and the shear layer have a pronounced effect on whether or not jets are stable against CDI, which may reconcile the results of global simulations conducted by \citet{2009MNRAS.394L.126M} and \citet{2010MNRAS.402....7M}, for example.
We will explore the interactions between the models we have presented here and a shear layer in a future work.

\section*{Acknowledgements}
We thank the anonymous referee for their helpful comments.
We acknowledge the support of NASA ATP grant NNX09AG02G and NSF grant AST-0907872.  This research was also supported in part by the National Science Foundation through XSEDE resources at the Texas Advanced Computing Center (TG-AST090106).  This work also utilized the Janus supercomputer, which is supported by the National Science Foundation (award number CNS-0821794) and the University of Colorado Boulder.  The Janus supercomputer is a joint effort of the University of Colorado Boulder, the University of Colorado Denver and the National Center for Atmospheric Research.  Visualization utilized VisIt, a visualization suite developed and maintained by Lawrence Livermore National Laboratory.

\bibliography{refs}

\label{lastpage}

\end{document}